\documentclass[sigconf]{acmart}

\usepackage{acmart-taps}
\usepackage{booktabs}
\usepackage{colortbl}
\usepackage{graphicx}
\usepackage{array}
\usepackage{booktabs}
\usepackage{pifont}
\usepackage{tikz}
\usetikzlibrary{automata, positioning}
\usepackage{soul}
\usepackage{xspace}
\usepackage{nicematrix}
\usepackage[english]{babel}
\usepackage{amsthm}
\usepackage{longtable}
\usepackage{multirow}
\usepackage{color}
\usepackage{listings}
\usepackage{wasysym}
\usepackage{multirow}
\usepackage{stfloats}
\usepackage{tabularx}
\usepackage{makecell}
\usepackage{amsmath,amsfonts}
\usepackage{algorithmic}
\usepackage{textcomp}
\usepackage{subcaption}
\usetikzlibrary{trees, shadows}
\usepackage{enumitem}
\usepackage{forest}

\usepackage[capitalise]{cleveref}
\crefformat{section}{\S#2#1#3}
\crefname{figure}{Figure}{Figures}
\crefname{appendix}{Appendix}{Appendices}
\crefname{table}{Table}{Tables}
\crefname{algorithm}{Algorithm}{Algorithms}
\crefname{listing}{Listing}{Listings}
\crefname{theorem}{Theorem}{Theorems}
\crefname{thm}{Theorem}{Theorems}
\crefname{lemma}{Lemma}{Lemmata}
\crefname{equation}{Eqt.}{Eqts.}
\crefformat{Grammar}{Grammar #1}

\setitemize{leftmargin=*}

\AtBeginDocument{%
  }

\theoremstyle{definition}

\newtheorem*{definition*}{Definition}

\newcommand{\etal}{\textit{et al.}\xspace}
\newcommand{\etals}{\textit{et al.}'s\xspace}

\newcommand{\eg}{{\em e.g.},\xspace}
\newcommand{\ie}{{\em i.e.},\xspace}
\newcommand{\et}{ {\em et al.}\xspace}

\newcommand{\vs}{{\em vs.}\xspace}

\definecolor{table_gray}{RGB}{235,235,235}

\newcommand{\numberofpaper}{{35}\xspace}

\newcommand{\numberofpatternpaper}{{3}\xspace}
\newcommand{\numberofstaticpaper}{{6}\xspace}
\newcommand{\numberofdynamicpaper}{{4}\xspace}
\newcommand{\numberofhybridpaper}{{6}\xspace}

\newcommand{\numberofMLdetectionpaper}{{1}\xspace}
\newcommand{\numberofregexmatchinpaper}{{3}\xspace}
\newcommand{\numberofresourcepaper}{{1}\xspace}
\newcommand{\numberofMLpaper}{{2}\xspace}
\newcommand{\numberofrepairpaper}{{4}\xspace}

\newcommand{\myparagraph}[1]{\textbf{#1}}

\newcommand{\subpara}[1]{\textbf{#1}}
\newcommand{\papersearchdate}{{June 2024}\xspace}

\definecolor{MyForestGreen}{RGB}{34,139,34}
\definecolor{MyMaroon}{RGB}{128,0,0}

\usepackage{pifont}
\newcommand{\cmark}{{\color{MyForestGreen}\ding{51}}}%
\newcommand{\xmark}{{\color{MyMaroon}\ding{55}}}%

\definecolor{editorGray}{rgb}{0.95, 0.95, 0.95}
\definecolor{editorOcher}{rgb}{1, 0.5, 0} 
\definecolor{editorGreen}{rgb}{0, 0.5, 0} 
\definecolor{btq}{rgb}{0.03, 0.91, 0.87} 
\definecolor{dtq}{rgb}{0.0, 0.81, 0.82} 
\definecolor{cdb}{rgb}{0.37, 0.62, 0.63} 

\lstdefinelanguage{js}{
	keywords={typeof, new, true, false, try, catch, function, return, null, catch, switch, var, if, in, for, while, do, else, case, break,let, const, throw, with},
	keywordstyle=\color{blue}\bfseries,
	ndkeywords={class, export, boolean, throw, implements, import, this, test, expect},
	ndkeywordstyle=\color{darkgray}\bfseries,
	identifierstyle=\color{black},
	sensitive=false,
	comment=[l]{//},
	morecomment=[s]{/*}{*/},
	commentstyle=\color{cyan}\ttfamily,
	stringstyle=\color{darkgray}\ttfamily,
	escapeinside={/*\#}{\#*/},
	morestring=[b]',
	morestring=[b]",
	morestring=[b]`
}

\definecolor{grassgreen}{rgb}{0.0, 0.5, 0.0}
\newif\ifcomments
\commentstrue
\commentsfalse
\ifcomments

  \newcommand{\JD}[1]{\textcolor{purple}{[Jamie says: #1]}}
  \newcommand{\BC}[1]{\textcolor{blue}{[Berk says: #1]}}
  \newcommand{\crs}[1]{[{\color{brown}Cris says: #1}]}
  \newcommand{\msd}[1]{\textcolor{grassgreen}{[Masud says: #1]}}
  \newcommand{\eat}[1]{}
  \newcommand{\EB}[1]{\textcolor{orange}{[Ethan says: #1]}}

\else

  \newcommand{\JD}[1]{}
  \newcommand{\BC}[1]{}
  \newcommand{\crs}[1]{}
  \newcommand{\msd}[1]{}
  \newcommand{\EB}[1]{}

\fi

\newif\ifANONYMOUS
\ANONYMOUSfalse

\acmConference[ASIA CCS '25]{ACM Asia Conference on Computer and Communications Security}{August 25--29, 2025}{Hanoi, Vietnam}

\acmBooktitle{ACM Asia Conference on Computer and Communications Security (ASIA CCS '25), August 25--29, 2025, Hanoi, Vietnam}

\setcopyright{cc}
\setcctype{by}
\copyrightyear{2025}
\acmYear{2025}
\acmDOI{10.1145/3708821.3733912}
\acmISBN{979-8-4007-1410-8/2025/08}

 \newcolumntype{g}{>{\columncolor{table_gray!50}}c}
\newcolumntype{G}{>{\columncolor{table_gray!50}}c}
\newcolumntype{H}{>{\columncolor{table_gray!50}}c}

\begin{document}

\title{SoK: A Literature and Engineering Review of Regular Expression Denial of Service (ReDoS)}

\author{Masudul Hasan Masud Bhuiyan}
\authornote{Both authors contributed equally to this research.}
\affiliation{%
 \institution{CISPA Helmholtz Center for Information Security}
 \city{Saarbrücken}
 \country{Germany}
}
\email{masudul.bhuiyan@cispa.de}

\author{Berk Çakar}
\authornotemark[1]
\affiliation{%
  \institution{Electrical and Computer Engineering\\Purdue University}
  \city{West Lafayette}
  \state{IN}
  \country{USA}
}
\email{bcakar@purdue.edu}

\author{Ethan H. Burmane}
\affiliation{%
\institution{Electrical and Computer Engineering\\Purdue University}
  \city{West Lafayette}
  \state{IN}
  \country{USA}
}
\email{eburmane@purdue.edu}

\author{James C. Davis}
\affiliation{%
\institution{Electrical and Computer Engineering\\Purdue University}
  \city{West Lafayette}
  \state{IN}
  \country{USA}
}
\email{davisjam@purdue.edu}

\author{Cristian-Alexandru Staicu}
\affiliation{%
  \institution{CISPA Helmholtz Center for Information Security}
  \city{Saarbrücken}
  \country{Germany}
}
\email{staicu@cispa.de}

\renewcommand{\shortauthors}{Bhuiyan, Çakar, Burmane, Davis, and Staicu}

\begin{abstract}
\looseness-1Regular Expression Denial of Service (ReDoS) is a vulnerability class that has become prominent in recent years.
Attackers can weaponize such weaknesses as part of asymmetric cyberattacks that exploit the slow worst-case matching time of regular expression (regex) engines.
In the past, problematic regexes have led to outages at Cloudflare and Stack Overflow, showing the severity of the problem.
While ReDoS has drawn significant research attention, there has been no systematization of knowledge to delineate the state of the art and identify opportunities for further research.
In this paper, we describe the existing knowledge on ReDoS.
We first provide a systematic literature review, discussing approaches for detecting, preventing, and mitigating ReDoS vulnerabilities.
Then, our engineering review surveys the latest regex engines to examine whether and how ReDoS defenses have been realized.
Combining our findings, we observe that
  (1) in the literature, almost no studies evaluate whether and how ReDoS vulnerabilities can be weaponized against real systems, making it difficult to assess their real-world impact;
  and
  (2) from an engineering view, many mainstream regex engines have introduced partial or full ReDoS defenses, rendering many threat models obsolete.
We conclude by highlighting avenues for future work.
The open challenges in ReDoS research are to evaluate emerging defenses and support engineers in migrating to defended engines.
We also highlight the parallel between performance bugs and asymmetric DoS, and we argue that future work should capitalize more on this similarity and adopt a more systematic view on ReDoS-like vulnerabilities.\vspace{-9pt}
\end{abstract}

\ccsdesc[300]{General and reference~Surveys and overviews}
\ccsdesc[300]{Security and privacy~Denial-of-service attacks}
\ccsdesc[300]{Software and its engineering~Empirical software validation}

\keywords{Systematization of knowledge (SoK), regular expression denial of service (ReDoS), regex engines, ReDoS defenses}

\maketitle

\section{Introduction} \label{sec:introduction}



Regular Expression Denial of Service (ReDoS) \cite{Crosby2003REDOSTalk, goyvaerts2003runaway, roichman2009redos} is a type of denial of service (DoS) attack \cite{NoteDenialofServiceProblem1983, noauthor_denial_nodate} that exploits inefficiencies in regular expression (regex) engines.
Many regex engines have worst-case exponential backtracking behavior.
By crafting input strings that trigger this behavior, attackers can cause systems to consume disproportionate CPU resources, leading to service disruptions.
ReDoS threatens real systems:
  slow regex matches caused outages at Stack Overflow~\cite{stackoverflow_outage} and Cloudflare~\cite{cloudflare_outage}.
ReDoS is the fourth most common server-side vulnerability class in JavaScript's npm ecosystem---only path traversal, prototype pollution, and command injection appear more often~\cite{bhuiyan_secbenchjs_2023}. 
It is also among the fastest-growing vulnerabilities in the same ecosystem \cite{snyk_report_redos}.
Consequently, there has been much research on ReDoS, with dozens of papers since 2015 (\S\ref{sec:litstudy}).
However, as yet there has been no systematization of knowledge about ReDoS, making it challenging to consolidate existing knowledge, identify gaps, and guide future efforts to address this complex and evolving threat.

To address this gap, we conducted a comprehensive literature review of recent papers on ReDoS (\S\ref{sec:litstudy}).
Our goals were to
  provide a tutorial on regexes and ReDoS,
  summarize previous work,
  assess progress in the field,
  and
  identify future work opportunities.
We analyze \numberofpaper papers on ReDoS vulnerabilities from top security conferences, focusing on detection, prevention, and mitigation strategies, to provide a comprehensive overview of the current state of the field.
One primary observation from our review was that almost all prior works rely on at least one of two main models of regex engine behavior:
  the first introduced in 1968 by Thompson~\cite{thompsonProgrammingTechniquesRegular1968},
  and
  the second in 1994 by Spencer~\cite{spencerRegularexpressionMatcher1994}.
While these models have been instrumental in shaping foundational research, they do not fully reflect modern regex engines, many of which now incorporate defenses against ReDoS attacks.
ReDoS research must be situated within real regex engine implementations, motivating us to perform an updated engineering review.

Thus, we conducted an engineering review of modern regex engines to address this disconnect between academic research and practical applications and update the understanding of the capabilities of regex engines (\S\ref{sec:engstudy}).
This review examines the implementations in the major programming languages to evaluate
  how they handle ReDoS vulnerabilities,
    either through
      new algorithms,
      resource caps,
      or other mitigations.
By analyzing these engines' designs, defenses, and real-world adoption, we provide a detailed account of the current state of regex processing and its implications for ReDoS research.
This engineering perspective complements the findings of our literature review and highlights the differences between traditional regex engines and contemporary regex engine behavior, offering up-to-date insights for future studies.

As a result of our synthesis of academic knowledge and engineering analysis,
we discuss five main findings (\S\ref{sec:discussion}).
In our review of the academic literature, we identified areas where threat models, ReDoS definitions, and attack evaluations could benefit from further refinement and alignment.
In our engineering review of regex engines, we note that four major engines have used three distinct algorithms to mitigate ReDoS, but these defenses do not always reduce the time complexity to linear.
We encourage the research community to explore opportunities to improve conceptual clarity in future ReDoS research and to focus efforts on evaluating and advancing state-of-the-art ReDoS defenses.

\underline{Our primary contributions are:}
\begin{itemize}
    \item We analyze \numberofpaper papers on ReDoS vulnerabilities from top security conferences, critiquing their definitions, methodologies, and assumptions. By identifying gaps in the literature, we provide a foundation for more practical and impactful research.

    \item We examine the current posture of the major regex engines, highlighting recent improvements, persistent vulnerabilities, and the trade-offs involved in mitigating ReDoS.

    \item Based on our findings, we propose guidelines for researchers to align their work with real-world needs and for practitioners to better evaluate and mitigate ReDoS risks.
\end{itemize}

\section{Background and Motivation} \label{sec:background}



In this section, we describe regexes, regex engines, and ReDoS.

\subsection{Regexes} \label{subsec:regexes}

\subsubsection{Kleene's regexes (K-regexes)}
Regexes are a popular technology for string matching tasks \cite{friedl2006mastering}.
Regexes are used in two basic operations:
  recognition (``Does this regex match this string?'')
  and
  parsing (``Extract relevant substrings from this match'')~\cite{hopcroftIntroductionAutomataTheory2007}.

Regexes began with Kleene's \cite{kleeneRepresentationEventsNerve1956} ``regular''  notation (\textit{K-regexes}), which specified a language as a set of strings using core constructs of concatenation (\texttt{.}), disjunction (\texttt{|}), and unbounded repetition (\texttt{*}). These operations define whether a given string belongs to the set of strings described by the regex.
Formally, a K-regex $R$ is:

\[
R \to \emptyset \;\Big{|}\; \epsilon \;\Big{|}\; \sigma \;\Big{|}\; R* \;\Big{|}\; R_1 \mid R_2 \;\Big{|}\; R_1 \cdot R_2
\]

\noindent
where
  the empty language is $\emptyset$,
  the empty string is $\epsilon$,
  and
  the alphabet is $\Sigma$.
A regex's semantics are defined by its \textit{language}, the set of strings it describes.
The language function \( L : R \to 2^{\Sigma^*} \) is:

\begin{table}[h]
\centering
{
\normalsize
\begin{tabular}{ll}
$L(\phi) = \phi$                &
$L(R\ast) = L(R)\ast$  \\
$L(\epsilon) = \{\epsilon\}$    &
$L(R_1|R_2) = L(R_1) \cup L(R_2)$   \\
$L(\sigma) = \{\sigma\}$        &
$L(R_1{\cdot}R_2) = L(R_1){\cdot}L(R_2)$
\end{tabular}
}
\end{table}

K-regexes can be modeled with nondeterministic or deterministic finite automata (NFA and DFA)~\cite{thompsonProgrammingTechniquesRegular1968,mcnaughtonRegularExpressionsState1960}.
These automata are 5-tuples $\langle Q, q_0, F, \Sigma, \delta \rangle$, where $Q$ is the set of states, $q_0$ is the start state, $F$ is the set of accepting states, $\Sigma$ is the input alphabet, and $\delta$ is the transition function.
Each regex operator---character matching, concatenation, repetition, and disjunction---has a corresponding NFA representation, as illustrated in Figure~\ref{fig:regex2NFA}.
These NFAs can be further converted into DFAs, providing deterministic evaluation at the cost of potentially higher state complexity~\cite{rabinFiniteAutomataTheir1959}.

\begin{figure}[ht]
    \centering
    \begin{minipage}[b]{0.23\columnwidth}
      \centering
      \begin{tikzpicture}[baseline=-2.5em,initial text=]
        \tikzstyle{every state}=[inner sep=0pt, minimum size=0.4cm, font=\scriptsize]

        \node[state, initial, initial distance=0.25cm]     (q0) {\scriptsize $q_0$};
        \node[state, right = 0.4cm of q0, accepting] (q1) {\scriptsize $q_1$};
        \draw [->] (q0) edge node [above] {\scriptsize $\sigma$} (q1);
      \end{tikzpicture}
      \caption*{\hspace{0.25cm}\small$\sigma$}
    \end{minipage}
    \begin{minipage}[b]{0.23\columnwidth}
      \centering
      \begin{tikzpicture}[baseline=-2.5em,initial text=]
        \tikzstyle{every state}=[inner sep=0pt, minimum size=0.4cm, font=\scriptsize]

        \node[state, initial, accepting, initial distance=0.25cm]     (q0) {\scriptsize $q_0$};
        \node[state, right = 0.4cm of q0, accepting] (q1) {\scriptsize $q_1$};
        \draw (q1) edge[loop above] node [right] {\scriptsize $L(R)$} (q1);
        \draw [->] (q0) edge node [above, yshift=1pt] {\scriptsize $L(R)$} (q1);
      \end{tikzpicture}
      \caption*{\hspace{0.25cm}\small$R\ast$}
    \end{minipage}
    \begin{minipage}[b]{0.23\columnwidth}
      \centering
      \begin{tikzpicture}[baseline=0em,initial text=]
        \tikzstyle{every state}=[inner sep=0pt, minimum size=0.4cm,font=\scriptsize]

        \node[state, initial, initial distance=0.25cm]         (q0) {\scriptsize $q_0$};
        \node[state, above right = 0.4cm of q0, accepting]     (q1) {\scriptsize $q_1$};
        \node[state, below right = 0.4cm of q0, accepting]     (q2) {\scriptsize $q_2$};
        \draw [->] (q0) edge node [above, pos=0.25, xshift=-6pt] {\scriptsize $L(R_1)$} (q1);
        \draw [->] (q0) edge node [below, pos=0.25, xshift=-6pt] {\scriptsize $L(R_2)$} (q2);
      \end{tikzpicture}
      \caption*{\hspace{0.25cm}\small$R_1 \mid R_2$}
    \end{minipage}
    \begin{minipage}[b]{0.23\columnwidth}
      \centering
      \begin{tikzpicture}[baseline=-2.5em,initial text=]
        \tikzstyle{every state}=[inner sep=0pt, minimum size=0.4cm, font=\scriptsize]

        \node[state, initial, initial distance=0.25cm]     (q0) {\scriptsize $q_0$};
        \node[state, above = 0.3cm of q0] (q1) {\scriptsize $q_1$};
        \node[state, right = 0.3cm of q0, accepting] (q2) {\scriptsize $q_2$};
        \draw [->] (q0) edge node [left] {\scriptsize $L(R_1)$} (q1);
        \draw [->] (q1) edge node [above, pos=0.75, xshift=6pt] {\scriptsize $L(R_2)$} (q2);
      \end{tikzpicture}
      \caption*{\hspace{0.25cm}\small$R_1 \cdot R_2$}
    \end{minipage}
    \hspace{0.02\columnwidth}
    \caption{
        K-regex operators and NFA equivalents, following the Thompson-McNaughton-Yamada construction~\cite{thompsonProgrammingTechniquesRegular1968,mcnaughtonRegularExpressionsState1960}.
    }
    \label{fig:regex2NFA}
\end{figure}

\subsubsection{Extended regexes (E-regexes)}
Over time, to simplify the representation, the K-regexes were ``extended'' (\textit{E-regexes}) with additional operators and constructs.
\textit{Syntax sugar} notations, like character ranges (\eg\texttt{[A-Z]} for \texttt{A|B|...|Z}) and one-or-more-repetition (\texttt{R+}), are still regular and can be transformed into equivalent K-regexes.
Some more complex extensions, such as atomic groups~\cite{berglundSemanticsAtomicSubgroups2017} and lookarounds (\eg~\cite{chattopadhyay2025verified,berglundRegularExpressionsLookahead2021,fujinamiEfficientMatchingMemoization2024}), have also been shown to be regular.
\JD{Let's clean this up with the appropriate references to some of the regular/not-regular literature. There are at least (1) a paper about atomic groups [Berglund], and (2) a paper about zero-width assertions. Check the papers from Rice, Stellenbosch, and ``The Japanese Guys''.}

Many other extensions have been made~\cite{friedl2006mastering,Hazel1997PCRE}, including irregular features such as backreferences~\cite{ahoPatternMatchingStrings1980}.
These extensions increase expressive power but exceed the boundaries of regular languages (\ie they cannot be directly modeled with finite automata), complicating theoretical analysis, implementation, and matching performance.
\JD{Give a cref to the table, not the appendix.}
A curated set of regex features is given in~Appendix~\ref{sec:appendix-regexSyntaxSemantics}.


\subsection{Regex Engines}\label{subsec:regexEnginesBackground}


Regexes are supported in all mainstream programming languages~\cite{friedl2006mastering} and are part of many other systems (\eg intrusion detection and prevention systems~\cite{roeschSnortLightweightIntrusion1999, RegexSnortRule}, web application firewalls~\cite{WafRegexPatternSet}, and logging tools~\cite{buchananRegexParsingExtract2023,AccessingEventData}).
These systems all implement or integrate a component called a \textit{regex engine} that provides support for the regex tasks of recognition and parsing.
As summarized by Davis~\cite{davis2020impact}, typical regex engine implementations followed one of two algorithms.
The first is a depth-first, backtracking search based on a library by Spencer~\cite{spencerRegularexpressionMatcher1994}.
The second is a breadth-first, lockstep search proposed by Thompson~\cite{thompsonProgrammingTechniquesRegular1968}.

The characteristics of these engine styles are summarized in Table~\ref{table:SpencerThompsonOverview}.
For the simple case of K-regexes, both can be modeled as NFA simulations.\footnote{Implementations of Thompson's algorithm typically have a direct mapping to an NFA simulation, while implementations of Spencer's algorithm may be implemented as recursive descent parsers. Here, we treat K-regexes, where both amount to NFA simulations.}
These approaches differ in their handling of \textit{ambiguity}~\cite{allauzen2008general}---where multiple transitions in the NFA are possible.
\begin{itemize}
\item Thompson's algorithm resolves ambiguity by tracking the set of all states in which the automata might be, updating this set using the transition function $\delta$ on each new character.
\item Spencer's algorithm resolves ambiguity by trying one path, and backtracking to try the other(s) if the first path fails.
\end{itemize}

\setcounter{table}{0}
{
\renewcommand{\arraystretch}{1.25} 
\begin{table}
    \footnotesize
    \centering
    \caption{
    Tradeoffs of Spencer and Thompson algorithms for regex engines~\cite{coxRegularExpressionMatching2007,davis2020impact}.
    Thompson's algorithm uses an efficient algorithm over an automata representation to obtain linear time complexity (in input string length),
    but this formal approach reduces its expressiveness.
    Spencer's algorithm gains broader E-regex support at the cost of exponential match times.
    }

    \begin{tabular}{cccc}
    \toprule
    \textbf{Algorithm} & \textbf{Strategy} & \textbf{Match time} & \textbf{Expressiveness}         \\
    \midrule
    Thompson's           & BFS NFA sim.    & Linear match & Limited E-regex support \\
    Spencer's           & Backtracking    & Exponential match & Easy E-regex support \\
    \bottomrule
    \end{tabular}
    \label{table:SpencerThompsonOverview}
\end{table}
}

Most of the time (for most regexes, inputs, and application contexts), both approaches run quickly enough and do not form a bottleneck for the broader application.
However, while the Thompson's algorithm approach guarantees match times that are linear in the input length, Spencer's algorithm exhibits worst-case polynomial or exponential match times on certain regex-input combinations.
Figure~\ref{fig:REDOSRegexes} gives two examples where Spencer's algorithm exhibits polynomial and exponential time complexity, respectively.
In each case, on an input $w=aa...ab$, the regex engine encounters ambiguity when processing an ``a'' symbol.
At the ultimate mismatch (caused by the final ``b''), the backtracking will take $|w|^2$ time for example (a) and $2^{|w|}$ time for example (b).
This worst-case behavior is colloquially termed \textit{catastrophic backtracking}. 
Note that the actual cost of the match depends on the length of the input; the term \textit{pumping} refers to the repetition of the problematic substring of the input (each ``a'' in this example).

\begin{figure}[ht]
    \centering
    \begin{minipage}[b]{0.45\columnwidth}
      \centering
      \begin{tikzpicture}[baseline=-1.65em,initial text=] 
        \tikzstyle{every state}=[inner sep=0pt, minimum size=0.4cm, font=\scriptsize]

        \node[state, initial, initial distance=0.25cm]     (q0) {\scriptsize $q_0$};
        \node[state, right = 0.4cm of q0, accepting] (q1) {\scriptsize $q_1$};
        \draw [->] (q0) edge node [above] {\scriptsize $\epsilon$} (q1);
        \draw [->] (q0) edge [loop above] node [right, pos=0.75] {\scriptsize $a$} (q0);
        \draw [->] (q1) edge [loop above] node [right, pos=0.75] {\scriptsize $a$} (q1);
      \end{tikzpicture}
      \caption*{\hspace{0.25cm}\small{(a) }NFA for $a{\ast}a{\ast}$}
    \end{minipage}
    \begin{minipage}[b]{0.45\columnwidth}
      \centering
      \begin{tikzpicture}[baseline=0em,initial text=]
        \tikzstyle{every state}=[inner sep=0pt, minimum size=0.4cm, font=\scriptsize]

        \node[state, initial, initial distance=0.25cm, accepting]     (q0) {\scriptsize $q_0$};
        \node[state, right = 0.4cm of q0, accepting] (q1) {\scriptsize $q_1$};
        \draw [->] (q0) edge [bend left, above] node {\scriptsize $a$ } (q1);
        \node[state, right = 0.4cm of q1, accepting] (q2) {\scriptsize $q_2$};
        \draw [->] (q1) edge node [above] {\scriptsize $\epsilon$ } (q2);
        \draw [->] (q1) edge [loop above] node [right, pos=0.75] {\scriptsize $a$} (q1);
        \draw [->] (q1) edge [bend left, below] node {\scriptsize $a$} (q0);
      \end{tikzpicture}
      \caption*{\hspace{0.25cm}\small{(b) }NFA for $(aa{\ast}){\ast}$}
    \end{minipage}
    \hspace{0.02\columnwidth}
    \caption{
       Example regexes that, in Spencer's algorithm, exhibit time complexity that is either: (a) polynomial or (b) exponential; in the length of the input string \boldmath$w=aa...ab$.
    }
    \label{fig:REDOSRegexes}
\end{figure}

Historically, regex engine maintainers accepted this worst-case behavior as the cost of supporting all E-regex features.
These extended features are relatively easy to implement in a (Spencer-style) backtracking engine, which often takes the form of a recursive descent parser~\cite{davis_using_2021}.
New E-regex features can be added by extending the set of parse operators, or equivalently by extending the meaning of the automaton's edge definitions ~\cite{davis2020impact}.
For example, zero-width assertions like \texttt{\textbackslash b} can use specialized $\epsilon$-edges that depend on passing a contextual test.
Backreferences can introduce edges that validate group content while advancing the input position accordingly.
Similarly, lookaround assertions can apply recursion to evaluate sub-patterns within the backtracking framework~\cite{davis_using_2021}.
These implementations are possible because Spencer's algorithm processes paths sequentially, which simplifies keeping track of path-specific behavior.



\subsection{Regular Expression Denial of Service (ReDoS)}\label{sec:redosBackground}

\subsubsection{ReDoS in Principle}
\label{subsec:redos-definition}

Regex engines often favor expressive power over predictable worst-case efficiency, and that engineering trade-off leaves them open to ReDoS~\cite{Crosby2003REDOSTalk}.
The Common Weakness Enumeration catalog labels the problem CWE-1333: \textit{Inefficient Regular Expression Complexity}~\cite{mitre_cwe1333}.
ReDoS is an example of an algorithmic complexity attack~\cite{crosby2003denial}, characterized by the small cost to create a malicious input and the large cost to process it.
Algorithmic complexity attacks are one kind of asymmetric DoS attacks~\cite{mantasAsymmetricDOS2015}, where the asymmetry comes from the worst-case performance of the data structures and algorithms used to implement the system.

Following Davis \etals\cite{davis_using_2021} definition, the threat model for a ReDoS attack has three ingredients:

\begin{enumerate}
  \item \textit{A Backtracking Regex Engine:} The regex engine employed for pattern matching in the victim system utilizes a backtracking approach (\S\ref{subsec:regexEnginesBackground}).
  Many regex engines implemented in programming languages (\eg JavaScript, Java, Python) have at some point met this criterion (\S\ref{sec:engstudy}). 
  \item \textit{A Vulnerable Regex:} The victim system uses a \textit{super-linear}, \textit{problematically ambiguous regex};
    for which the attacker crafts \textit{attack strings} that take polynomial or exponential time to match in the string length~\cite{wustholz_static_2017,weideman_analyzing_2016}. 
  \item \textit{An Adversary-Controlled Input:} The attacker supplies inputs that result in attack strings reaching the vulnerable regex and triggering its worst-case time complexity. 
\end{enumerate}

Given these three ingredients, if the system has insufficient mitigations (safeguards) in place, then the attacker can cause the victim system to consume excessive computational resources (\eg CPU and memory).
The attacker may need to tailor the number of pumps based on the length of typical input (to avoid anomaly detection) or the maximum length allowed (to avoid truncation).



\begin{figure}
\centering
\includegraphics[width=\linewidth]{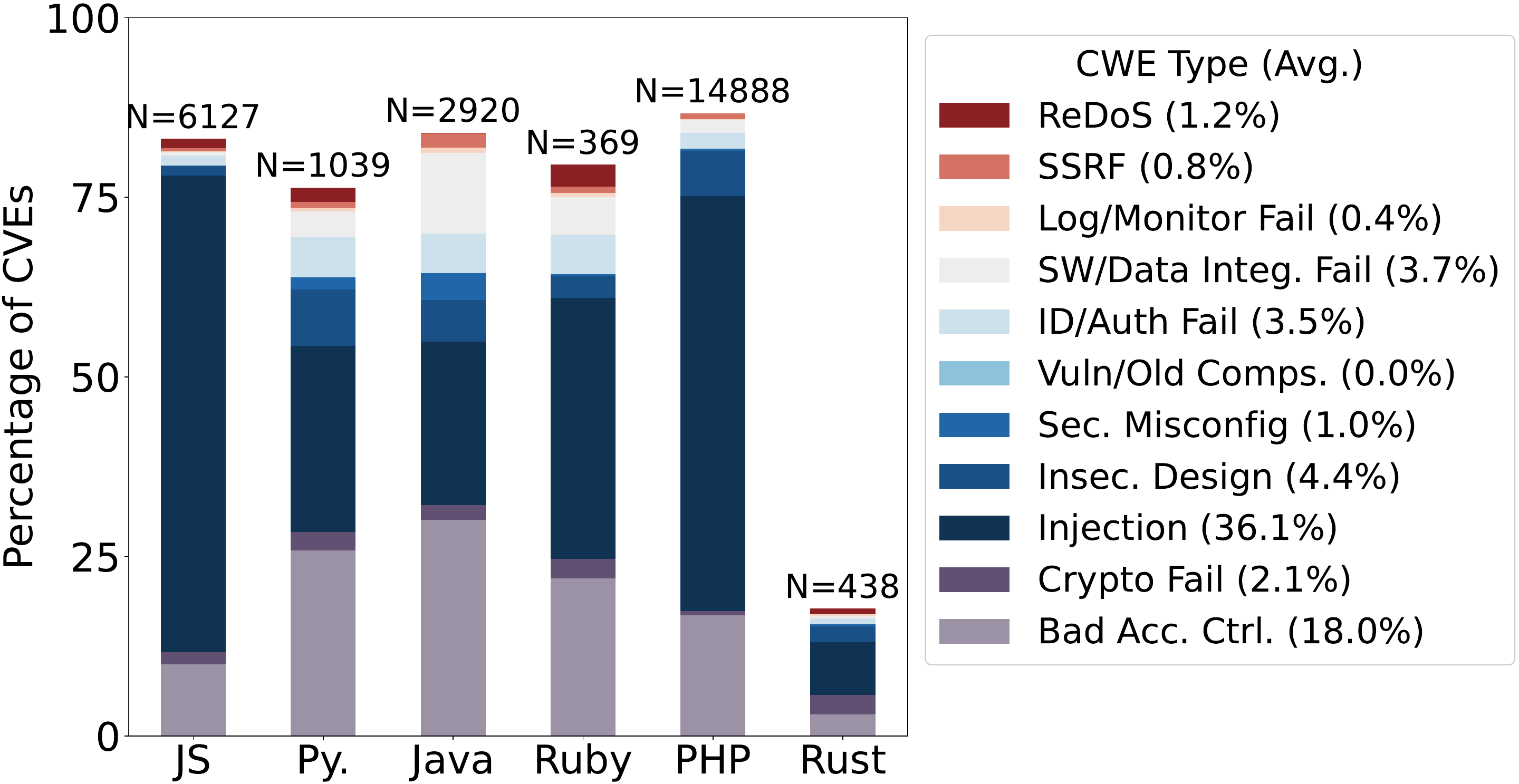}
\caption{
  Distribution of OWASP Top 10 and ReDoS CVEs per ecosystem, showing average prevalence per category.
  N indicates the total number of CVEs recorded for each ecosystem.
  The data cover 2014-2023.
  For details, see~Appendix~\ref{sec:appendix-REDOSInPractice}.
}
\label{fig:Motivation-CVEData}
\end{figure}

\subsubsection{ReDoS in Practice}
\label{subsec:redos-in-os}

In recent years, ReDoS has become a common security concern.
Two major web services had outages that resulted from slow regex behavior:
  Stack Overflow in 2016~\cite{stackoverflow_outage}
  and
  Cloudflare in 2019~\cite{cloudflare_outage}.
While these outages did not involve an adversary (\ie they were reliability issues, not security issues), they made service providers aware of the risk of ReDoS.
Since 2020, the security company Snyk has observed a jump in reported ReDoS vulnerabilities, especially in the JavaScript package ecosystem npm~\cite{snyk_report_redos}.
In parallel with this surge, a rich body of grey literature has emerged to offer practical guidance on repairing ReDoS vulnerabilities~\cite{snyk_redos,githubblog,davis_blog,regular_expressions_info_blog}.

For a sense of the formal disclosure of ReDoS vulnerabilities, we examined the National Vulnerability Database (NVD) for Common Vulnerabilities and Exposures (CVE) records pertaining to ReDoS.
Following the keyword method of Hassan \etal~\cite{hassan_improving_2023},
since 2014 $\sim$400 ReDoS CVEs have been disclosed in software from six major programming language ecosystems.
As a reference point, we compared these results to CVEs related to the OWASP Top 10~\cite{OWASPTop10}.
Averaging over 2014-2023, Figure~\ref{fig:Motivation-CVEData} shows that the rate of ReDoS exceeds four of the OWASP Top 10 vulnerability types.

\section{Academic Research on ReDoS}
\label{sec:litstudy}

Contemporaneous with the increasing industry recognition of ReDoS, ReDoS has become a common subject of cybersecurity research.
However, there has been no systematization of the resulting knowledge.
We therefore conducted a systematic literature review of ReDoS papers published between 2015 and 2024 at leading conferences in computer security (S\&P, USENIX, CCS, NDSS), software engineering (ICSE, FSE, ASE), and programming languages (PLDI, POPL, OOPSLA). Following the structured methodology outlined by Schloegel \etal~\cite{schloegel2024sok}, we identified relevant papers in conference proceedings via an initial keyword search conducted in \papersearchdate, using the terms  \texttt{ReDoS}, \texttt{denial-of-service}, \texttt{regex}, \texttt{regular expression}, \texttt{regular expression analysis}, and \texttt{regular expression execution}. This yielded 93 papers, which were further filtered via the following criteria:

\begin{itemize}
    \item \textbf{Inclusion Criteria:} Papers that primarily address ReDoS, propose new attacks or defenses against ReDoS, or extensively study existing ReDoS attacks or defenses.
    \item \textbf{Exclusion Criteria:} Papers that focus on general DoS or those that discuss regexes, without specifically addressing ReDoS. Appendix~\ref{subsec:excluded-works} provides further discussion of the excluded works.
\end{itemize}

To ensure informed inclusion and exclusion decisions, we reviewed the abstract, introduction, and evaluation sections of each paper. We identified \numberofpaper{} papers that met our inclusion criteria and systematically analyzed them along four key dimensions:

\begin{enumerate}
    \item[(1)]\textbf{Research Directions}: Captures the primary focus of each study, which typically falls into one of three categories:
\end{enumerate}

    \begin{itemize}
        \item \textit{Detection}: Identifying and analyzing vulnerabilities.
        \item \textit{Prevention}: Addressing root causes by improving regex engines or limiting unsafe constructs.
        \item \textit{Mitigation}: Reducing attack impact through runtime defenses or repair mechanisms.
    \end{itemize}

\begin{enumerate}
    \item [(2)]\textbf{Threat Models}: Examines the assumptions about attacker capabilities (\eg control over input) and system behavior (\eg rate-limiting or regex engine design), highlighting the practical applicability of each study.
\end{enumerate}

\begin{enumerate}
    \item [(3)]\textbf{Vulnerability Definitions}: Focuses on how papers describe what constitutes a ReDoS vulnerability or a vulnerable regex (\eg having exponential matching time).
\end{enumerate}

\begin{enumerate}
    \item [(4)]\textbf{Evaluation Methods}: Assesses the languages and regex engines used for testing, providing insights into the platforms where these approaches are validated.
\end{enumerate}


Additionally, we reviewed other closely related works that, while not meeting our strict inclusion criteria, are still highly relevant to ReDoS. For instance, papers such as \cite{coxRegularExpressionMatching2007, holik_fast_2023, smith_xfa_2008} propose improved regex matching algorithms. Although mitigating ReDoS was not their primary focus, these works provide valuable insights and serve as foundational building blocks for ReDoS defenses, making them significant for academics and researchers in the field.

We begin by discussing the main existing research directions~(\S\ref{subsec:research-directions}), distinguishing between detection, prevention, mitigation, and other related works. We then explore the assumptions made about the attackers' capabilities or the underlying systems~(\S\ref{subsec:threat-models}), along with the definitions of ReDoS vulnerabilities or successful attacks provided in prior works~(\S\ref{subsec:redos-definitions}). Finally, we discuss how the studied papers were evaluated~(\S\ref{subsec-evaluation}).



\begin{figure*}[t]
\raggedright
\tikzset{
    my node/.style={
        draw=gray,
        thick,
        rounded corners=5pt,
        minimum width=1cm,
        text height=1.5ex,
        text depth=0ex,
        font=\sffamily,
    },
    detection/.style={
        inner color=blue!10,
        outer color=blue!20,
    },
    prevention/.style={
        inner color=green!10,
        outer color=green!20,
    },
    mitigation/.style={
        inner color=yellow!10,
        outer color=yellow!20,
    },
    related/.style={
        inner color=orange!10,
        outer color=orange!20,
    },
    subcategory/.style={
        inner color=gray!10,
        outer color=gray!20,
    }
}

\begin{forest}
for tree={
    my node,
    grow=east,
    l sep=15pt,
    s sep=5pt,
    edge={gray, thick},
    parent anchor=east,
    child anchor=west,
    edge path={
        \noexpand\path [draw, \forestoption{edge}]
        (!u.east) -- ++(10pt,0) |- (.west)\forestoption{edge label};
    }
}
[ReDoS Systematization
    [Other Related Work, related
        [LLM-based Studies, related
            [\cite{siddiq_regexdoseval_2024, siddiq_understanding_2024}, related]
        ]
        [Empirical Studies, related
            [\cite{staicu_freezing_nodate, davis_why_2019}, related]
        ]
        [Dataset Studies, related
            [\cite{bhuiyan_secbenchjs_2023}, related]
        ]
    ]
    [Mitigation, mitigation
        [Resource Limitation, mitigation
            [\cite{davis_sense_nodate}, mitigation]
        ]
        [Anomaly Detection, mitigation
            [\cite{bai_runtime_2021, tandon_leader_2023}, mitigation]
        ]
    ]
    [Prevention, prevention
        [Automatic Repair, prevention
            [\cite{li_regexscalpel_nodate, li_flashregex_nodate, chida_repairing_2022, hassan_improving_2023}, prevention]
        ]
        [Better Regex Matching Algorithms, prevention
            [\cite{davis_using_2021, davis_rethinking_2019, turonova_regex_2020}, prevention]
        ]
    ]
    [Detection, detection
        [Machine Learning (ML)-based Detection, detection
            [\cite{demoulin2019detecting}, detection]
        ]
        [Hybrid Detection, detection
            [\cite{wustholz_static_2017, liu_revealer_2021, li_redoshunter_nodate, noller_badger_2018, wang_effective_2023, davis_impact_2018}, detection]
        ]
        [Dynamic Detection, detection
            [\cite{shen_rescue_2018, mclaughlin_regulator_nodate, barlas_exploiting_2022, petsios_slowfuzz_2017}, detection]
        ]
        [Automata-based Static Analysis, detection
            [Enhanced Automata Models, detection
                [\cite{parolini_sound_2023}, detection]
            ]
            [DFA-based Approaches, detection
                [\cite{turonova_counting_nodate, su2024towards}, detection]
            ]
            [NFA-based Approaches, detection
                [\cite{rathnayake_static_2017, kirrage_static_2013, weideman_analyzing_2016}, detection]
            ]
        ]
        [Heuristic-based Detection, detection
            [\cite{davis_case_2017, kluban_measuring_2022, kluban_detecting_2024}, detection]
        ]
    ]
]
\end{forest}

\caption{Systematization of ReDoS research papers categorized into detection, prevention, mitigation, and related studies, with subcategories and key references.}
\label{fig:redos_systematization_compact}
\vspace{-3mm}
\end{figure*}


\subsection{Main Existing Research Directions}
\label{subsec:research-directions}

Figure~\ref{fig:redos_systematization_compact} provides an overview of the main research directions in the study of ReDoS vulnerabilities, categorized into detection, prevention, mitigation, and complementary studies. Detection focuses on identifying ReDoS vulnerabilities in regexes and understanding their exploitability. Prevention aims to address the root causes of ReDoS vulnerabilities by improving regex engines or restricting vulnerable constructs. Mitigation, on the other hand, focuses on minimizing the impact of ReDoS attacks by deploying runtime safeguards. We discuss these categories in detail below.

\subsubsection{Detection}

Detection is a critical area of research in ReDoS, with studies focusing on identifying vulnerable regexes across various ecosystems. Nine papers analyze open-source package ecosystems, such as npm~\cite{bhuiyan_secbenchjs_2023, staicu_freezing_nodate}, pip~\cite{davis_impact_2018, wang_effective_2023}, and Maven~\cite{wang_effective_2023, li_redoshunter_nodate}, while seven focus on testing live web applications~\cite{staicu_freezing_nodate, barlas_exploiting_2022}. Additionally, 27 papers investigate ReDoS vulnerabilities using curated regex datasets~\cite{turonova_counting_nodate, liu_revealer_2021, siddiq_understanding_2024, siddiq_regexdoseval_2024}. We categorize detection methods into five key approaches: \textit{Heuristic-based Detection} (three papers), \textit{Automata-based Detection} (seven papers), \textit{Dynamic Detection} (four papers), \textit{Hybrid Detection} (five papers), and \textit{Machine Learning (ML)-based Detection} (one paper). Heuristic-based techniques rely on simple pattern rules to flag problematic regexes, offering speed but often at the cost of high false positive rates. Automata-based approaches, which model regexes using finite-state machines, provide formal accuracy but are computationally expensive. Dynamic detection evaluates regex behavior on specific inputs, revealing real-world vulnerabilities but requiring significant runtime resources. Hybrid methods combine static and dynamic techniques to balance precision and efficiency, while emerging ML approaches use regex datasets to predict vulnerabilities, opening new avenues but facing challenges like adversarial examples and limited training data. This categorization highlights the trade-offs in speed, accuracy, and false positive/negative rates among different techniques, offering a structured view of existing detection strategies and their potential for future enhancement.

\myparagraph{Heuristic-based Detection (\numberofpatternpaper studies):} There are many approaches that employ simple pattern-matching techniques to identify potentially vulnerable regexes. Studies by Kluban \etal~\cite{kluban_detecting_2024, kluban_measuring_2022} and Davis \etal~\cite{davis_case_2017} utilize tools like regexploit~\cite{noauthor_doyensecregexploit_nodate}, redos-detector~\cite{jenkinson_tjenkinsonredos-detector_2024}, and safe-regex~\cite{noauthor_davisjamsafe-regex_nodate} to identify these vulnerabilities by searching for constructs such as infinite repeats (\texttt{a*b*a*}), branches ($a|b$), or nested quantifiers (\texttt{*}, \texttt{+}, \texttt{?}, \texttt{\{n,m\}}), which are often associated with ReDoS vulnerabilities. While these methods are fast, enabling the rapid analysis of large numbers of regexes, they typically focus on syntax rather than semantics, leading to potential accuracy issues. For instance, Parolini \etal~\cite{parolini_sound_2023} demonstrated that while tools like regexploit achieved low false positive and negative rates, others like safe-regex and redos-detector exhibited significantly higher false positive rates. This highlights the primary limitation of heuristic-based approaches, which often require further analysis to confirm vulnerabilities.\enlargethispage{12pt}

\myparagraph{Automata-based Static Analysis (\numberofstaticpaper studies):}

Automata-based static analysis is a powerful approach for detecting problematic regexes without requiring program execution. This method has been explored in several studies~\cite{rathnayake_static_2017, kirrage_static_2013, parolini_sound_2023, turonova_counting_nodate, su2024towards, weideman_analyzing_2016}, which employ automata models such as NFAs, DFAs, and enhanced automata to evaluate regex semantics and detect vulnerabilities that lead to ReDoS attacks.

\subpara{NFA-based methods}, such as those by Kirrage \etal~\cite{kirrage_static_2013} and Rathnayake \etal~\cite{rathnayake_static_2017}, construct NFAs to analyze patterns like \texttt{(prefix, pumpable string, suffix)} and identify vulnerabilities caused by exponential backtracking. Building on this, Weideman \etal~\cite{weideman_analyzing_2016} enhance NFAs with prioritization (pNFA) to reduce redundant state exploration, providing a more precise analysis of backtracking vulnerabilities.

\subpara{DFA-based approaches}, as demonstrated by Turoňová \etal~\cite{turonova_counting_nodate} and Shen \etal~\cite{su2024towards}, rely on deterministic automata simulations to evaluate vulnerabilities in non-backtracking engines. By incorporating specialized counting mechanisms, these approaches effectively detect problematic patterns, such as bounded quantifiers and repetitive constructs, while optimizing resource usage. This focus on determinism allows for efficient analysis of regexes without the ambiguity associated with backtracking.

\subpara{Enhanced automata models} offer a novel perspective on regex mathing. Parolini \etal~\cite{parolini_sound_2023} introduce tree semantics, extending beyond traditional NFA and DFA approaches by capturing the structural interactions of regex engines. This enables the detection of vulnerabilities that are difficult to represent in standard automata.

While these methods excel in early detection, they face challenges such as false positives from over-approximations, false negatives from under-approximations, and high computational complexity for analyzing intricate regex structures. Despite these trade-offs, automata-based static analysis remains a foundational tool for addressing ReDoS vulnerabilities.\enlargethispage{12pt}

\myparagraph{Dynamic Analysis (\numberofdynamicpaper studies):} Dynamic approaches~\cite{shen_rescue_2018, mclaughlin_regulator_nodate, barlas_exploiting_2022, petsios_slowfuzz_2017} to detecting ReDoS vulnerabilities involve executing regex\-es with various inputs and analyzing their runtime behavior to identify potential issues. For instance, Shen\et~\cite{shen_rescue_2018} developed ReScue, a tool that uses an evolutionary genetic algorithm to generate time-consuming strings capable of triggering ReDoS vulnerabilities. Similarly, Barlas \etal~\cite{barlas_exploiting_2022} and McLaughlin \etal~\cite{mclaughlin_regulator_nodate} utilized dynamic analysis with fuzzing techniques, while Petsios \etal~\cite{petsios_slowfuzz_2017} introduced SlowFuzz, which uses automated fuzzing to find inputs causing worst-case algorithmic behavior.

\myparagraph{Hybrid Approaches (\numberofhybridpaper studies):}
Existing static analysis methods for detecting ReDoS vulnerabilities often face a trade-off between precision and recall. For example, Davis \etal~\cite{davis_impact_2018} and Shen \etal~\cite{shen_rescue_2018} reported low F-1 scores of 44.94\% and 3.57\%, respectively~\cite{li_redoshunter_nodate}. To address these limitations, dynamic validation is frequently employed alongside static analysis, resulting in hybrid approaches. Tools such as NFAA~\cite{wustholz_static_2017}, Revealer~\cite{liu_revealer_2021}, ReDoSHunter~\cite{li_redoshunter_nodate}, Badger~\cite{noller_badger_2018}, and RENGAR~\cite{wang_effective_2023} combine static and dynamic methods to mitigate false positives. For instance, hybrid approaches like ReDoSHunter integrate static and dynamic techniques to improve detection accuracy, while tools such as NFAA refine dynamic testing with insights from static analysis~\cite{wustholz_static_2017}. These methods effectively balance the strengths of both static and dynamic detection, reducing the likelihood of inaccurate results.

\myparagraph{Machine Learning (ML)-based Detection (\numberofMLdetectionpaper study):}
ML techniques are also gaining traction in ReDoS detection. Unlike hybrid methods, which explicitly integrate static and dynamic analysis, ML approaches aim to predict vulnerabilities and detect anomalies based on patterns in data. For example, Demoulin \etal~\cite{demoulin2019detecting} introduced a semi-supervised learning model to analyze resource utilization and identify potential vulnerabilities. This method triggers alerts and provides mitigation strategies by detecting anomalies in resource consumption. While ML approaches offer promise, they face challenges such as reliance on labeled data, vulnerability to adversarial inputs, and difficulty in addressing context-specific patterns that may not generalize well across diverse regex use cases.\enlargethispage{12pt}

\subsubsection{Prevention}
Prevention strategies focus on addressing the root causes of ReDoS vulnerabilities, either by improving regex matching algorithms or by repairing vulnerable regex patterns. These approaches aim to prevent ReDoS attacks entirely by designing systems that eliminate worst-case performance scenarios, thus avoiding the need for runtime detection or mitigation.

\myparagraph{Better Regex Matching Algorithms (\numberofregexmatchinpaper studies):}
Researchers have proposed developing new regex matching engines to address ReDoS vulnerabilities at a fundamental level. Davis \etal~\cite{davis_using_2021, davis_rethinking_2019} introduced memoization-based optimization to speed up regex matching by caching intermediate results, significantly reducing redundant computations. Turoňová \etal~\cite{turonova_regex_2020, turonova_counting_nodate} proposed novel counting automata matching algorithms that further optimize counter management during matching to efficiently handle regexes with bounded repetitions. These new engines often employ finite-state machine-based approaches instead of backtracking, inherently avoiding the exponential time complexity that leads to ReDoS vulnerabilities. Additionally, some methods impose restrictions on regex features to ensure faster and safer processing.

\myparagraph{Automatic Repair (\numberofrepairpaper studies):}
Automatic repair techniques aim to proactively address vulnerabilities in regexes to mitigate ReDoS attacks. Several studies have explored programmatic frameworks to detect and repair problematic regex patterns. For example, Li \etal~\cite{li_flashregex_nodate} developed FlashRegex, a programming-by-example (PbE) tool that repairs or synthesizes regex from provided matching examples. However, FlashRegex struggles with complex regex features, such as lookarounds and backreferences~\cite{li_regexscalpel_nodate}. To address these limitations, RegexScalpel~\cite{li_regexscalpel_nodate} employs a localize-and-fix approach, which identifies fine-grained vulnerabilities and applies predefined repair strategies to improve recall and precision.
In a complementary effort, Hassan \etal~\cite{hassan_improving_2023} focus on identifying specific regex structures or ``anti-patterns'' that are prone to ReDoS vulnerabilities. Their method analyzes regex patterns to uncover and repair potential vulnerabilities early in the development process, avoiding the need for runtime execution. This approach complements programmatic frameworks by systematically identifying and addressing problematic constructs.

Other tools, such as Chida \etal~\cite{chida_repairing_2022}, also leverage PbE algorithms but face challenges related to dependency on user-provided examples, which can result in incomplete repairs~\cite{li_redoshunter_nodate, wang_effective_2023}. These methods, while effective for certain regex patterns, often struggle with advanced or highly customized regex constructs, underscoring the need for more comprehensive solutions capable of addressing complex vulnerabilities. Moreover, improving the generalization and robustness of automatic repair frameworks remains a key area of ongoing research.

Prevention techniques address ReDoS vulnerabilities by improving the efficiency and resilience of regex engines or by repairing vulnerable regex patterns. While they significantly reduce the risk of attacks, these methods require careful configuration to ensure compatibility with complex patterns and real-world scenarios.\enlargethispage{12pt}

\subsubsection{Mitigation}
Mitigation strategies aim to detect ReDoS attacks in live systems and mitigate their impact in real-time. These approaches focus on monitoring runtime behavior to identify anomalies in resource consumption, such as excessive memory usage or prolonged processing times, and applying defensive measures to minimize the effects of ongoing attacks. Unlike prevention, which aims to eliminate vulnerabilities before execution, mitigation provides a dynamic response to attacks as they occur.

\myparagraph{Anomaly Detection (\numberofMLpaper studies):}
Anomaly detection techniques are central to mitigation strategies, leveraging real-time monitoring to identify unusual patterns indicative of ReDoS attacks. ML-based approaches have proven effective in this domain~\cite{bai_runtime_2021, tandon_leader_2023}. For example, Tandon \etal~\cite{tandon_leader_2023} employ the Elliptic Envelope model to analyze runtime features such as request time, memory usage, and CPU cycles, enabling the system to detect deviations from normal behavior. Bai \etal~\cite{bai_runtime_2021} introduce a feedback loop mechanism that continuously adjusts defensive measures based on real-time performance metrics, providing robust protection against zero-day ReDoS attacks.

While promising, ML-based anomaly detection methods are not without challenges. For instance, Tandon \etal~\cite{tandon_leader_2023} report a false positive rate of 1.3\% and a false negative rate of 0.4\%, which could still impact practical deployment. Furthermore, these systems can be vulnerable to adversarial attacks, where carefully crafted inputs are designed to bypass detection mechanisms~\cite{bai_runtime_2021}. These limitations underscore the need for more resilient and adaptive approaches to anomaly detection, ensuring robustness against both known and emerging threats.

\myparagraph{Limiting Resource Consumption (\numberofresourcepaper study):}
To mitigate ReDoS attacks, Davis \etal~\cite{davis_sense_nodate} proposed using first-class timeouts as a defense against Event Handler Poisoning (EHP) in Node.js. Their prototype, \texttt{Node.cure}, effectively limits processing time with minimal performance overhead. Similarly, runtime environments like PHP, Perl, and .NET have adopted resource-limiting mechanisms to curb backtracking and excessive processing. In PHP, the pcre.backtrack\_limit and pcre.recursion\_limit directives control the number of backtracks and recursion depth during regex matching \cite{PHPRuntimeConfiguration}. Perl's PCRE2 library allows setting evaluation limits via the eval function, with a default recursion depth limit of 10M. In .NET, the RegexOptions.MatchTimeout property enables developers to specify a maximum regex processing duration. While these strategies effectively cap resources to reduce ReDoS risks, they are often too coarse-grained, making it challenging to balance performance and avoid false positives in complex regex patterns or large datasets.

\subsubsection{Other Related Work}

Beyond approaches for addressing ReDoS directly, several complementary studies provide valuable insights into datasets, real-world analyses, and novel frameworks.

\myparagraph{Dataset Studies (1 study):}
Bhuiyan \etal~\cite{bhuiyan_secbenchjs_2023} introduced Sec\-Bench\-JS, a benchmark dataset for evaluating the security of JavaScript applications, including regex vulnerabilities. Such datasets enable standardized testing and improve ReDoS detection tools.

\myparagraph{Empirical Studies (2 studies):}
Staicu \etal~\cite{staicu_freezing_nodate} analyzed 2846 popular websites, finding 339 vulnerable to ReDoS attacks, highlighting the risks of server-side JavaScript. Davis \etal~\cite{davis_why_2019} studied regex portability across eight programming languages, uncovering semantic and performance differences in 25\% of regexes. These studies underscore the need for safer regex practices and consistent regex engine implementations.

\myparagraph{LLM-based Studies (2 studies):}
Siddiq \etal~\cite{siddiq_regexdoseval_2024} proposed Re(gEx|DoS)Eval, a framework for evaluating LLM-generated regexes using metrics like pass@k and vulnerable@k, demonstrating its use on T5, Phi-1.5, and GPT-3.5-Turbo. In another study, Siddiq \etal~\cite{siddiq_understanding_2024} categorized vulnerable regex patterns and analyzed developer discussions, finding that GPT-3.5-Turbo often generates regexes with polynomial vulnerabilities.

These studies provide benchmarks, empirical data, and frameworks that complement prevention, detection, and mitigation strategies, paving the way for robust ReDoS solutions.

\subsection{Threat Models}
\label{subsec:threat-models}


\aptLtoX{\begin{table}[t]
\normalsize
\centering
\caption{
Overview of ReDoS threat models in prior work. Each column shows whether the study analyzes code fragments (standalone regex snippets), use within libraries, or full applications, and whether it assumes the attacker can send an arbitrary number of requests.
The symbol \CIRCLE~denotes the paper covers that aspect in detail; -- means it does not.
}
\label{tab:threat_model}
\resizebox{\linewidth}{!}{%
  \begin{tabular}{lcccc}
    \toprule
      \textbf{Paper}
    & \textbf{Code Fragment}
    & \textbf{Library}
    & \textbf{Application}
    & \begin{tabular}[c]{@{}l@{}}\textbf{Arbitrary Number}\\\textbf{of Requests}\end{tabular}  \\
    \midrule
    Su \etal~\cite{su2024towards}              & \CIRCLE & -- & -- & 100kB \\
    Davis \etal~\cite{davis_case_2017}         & \CIRCLE & -- & -- & -- \\
    Turoňová \etal~\cite{turonova_counting_nodate}
                                               & \CIRCLE & -- & -- & 1.5K–9K \\
    Liu \etal~\cite{liu_revealer_2021}         & \CIRCLE & -- & -- & 128 \\
    Staicu \etal~\cite{staicu_freezing_nodate} & --       & \CIRCLE & \CIRCLE & 82K \\
    Davis \etal~\cite{davis_impact_2018}      & \CIRCLE & \CIRCLE & -- & 100K–1M \\
    Siddiq \etal~\cite{siddiq_understanding_2024}
                                               & \CIRCLE & -- & -- & -- \\
    Shen \etal~\cite{shen_rescue_2018}         & \CIRCLE & -- & -- & 128 \\
    Barlas \etal~\cite{barlas_exploiting_2022} & --       & -- & \CIRCLE & -- \\
    Siddiq \etal~\cite{siddiq_regexdoseval_2024}
                                               & \CIRCLE & -- & -- & -- \\
    Davis \etal~\cite{davis_why_2019}          & \CIRCLE & -- & -- & -- \\
    Kluban \etal~\cite{kluban_measuring_2022}  & \CIRCLE & -- & -- & -- \\
    Bhuiyan \etal~\cite{bhuiyan_secbenchjs_2023}
                                               & \CIRCLE & \CIRCLE & -- & -- \\
    Noller \etal~\cite{noller_badger_2018}     & \CIRCLE & -- & -- & -- \\
    Wang \etal~\cite{wang_effective_2023}      & \CIRCLE & \CIRCLE & -- & -- \\
    Rathnayake \etal~\cite{rathnayake_static_2017}
                                               & \CIRCLE & -- & -- & -- \\
    McLaughlin \etal~\cite{mclaughlin_regulator_nodate}
                                               & \CIRCLE & \CIRCLE & -- & 1M \\
    Li \etal~\cite{li_redoshunter_nodate}     & \CIRCLE & \CIRCLE & -- & -- \\
    Parolini \etal~\cite{parolini_sound_2023}  & \CIRCLE & -- & -- & 128 \\
    Kirrage \etal~\cite{kirrage_static_2013}  & \CIRCLE & -- & -- & -- \\
    Wüstholz \etal~\cite{wustholz_static_2017}
                                               & \CIRCLE & -- & \CIRCLE & 140K \\
    Petsios \etal~\cite{petsios_slowfuzz_2017} & \CIRCLE & -- & -- & -- \\
    Kluban \etal~\cite{kluban_detecting_2024}  & --       & \CIRCLE & -- & -- \\
    Demoulin \etal~\cite{demoulin2019detecting}& --       & -- & \CIRCLE & -- \\
    Chida \etal~\cite{chida_repairing_2022}    & \CIRCLE & -- & -- & -- \\
    Bai \etal~\cite{bai_runtime_2021}          & --       & -- & \CIRCLE & 50K \\
    Li \etal~\cite{li_regexscalpel_nodate}     & \CIRCLE & \CIRCLE & -- & 1M \\
    Davis \etal~\cite{davis_rethinking_2019}   & \CIRCLE & -- & -- & -- \\
    Li \etal~\cite{li_flashregex_nodate}       & \CIRCLE & -- & -- & -- \\
    Hassan \etal~\cite{hassan_improving_2023}  & \CIRCLE & -- & -- & -- \\
    Davis \etal~\cite{davis_sense_nodate}      & --       & \CIRCLE & \CIRCLE & -- \\
    Davis \etal~\cite{davis_using_2021}        & \CIRCLE & -- & -- & 10K–100K \\
    Tandon \etal~\cite{tandon_leader_2023}     & --       & -- & \CIRCLE & -- \\
    Weideman \etal~\cite{weideman_analyzing_2016}
                                               & \CIRCLE & -- & -- & -- \\
    Turoňová \etal~\cite{turonova_regex_2020}  & \CIRCLE & -- & -- & 500K–50M \\
    \bottomrule
  \end{tabular}%
}
\end{table}
}{
\begin{table}[t]
\normalsize
\centering
\caption{
Overview of ReDoS threat models in prior work. Each column shows whether the study analyzes code fragments (standalone regex snippets), use within libraries, or full applications, and whether it assumes the attacker can send an arbitrary number of requests.
The symbol \CIRCLE~denotes the paper covers that aspect in detail; -- means it does not.
}
\label{tab:threat_model}
\resizebox{\linewidth}{!}{%
  \begin{tabular}{lgggg}
    \toprule
    \rowcolor{white}
      \textbf{Paper}
    & \textbf{Code Fragment}
    & \textbf{Library}
    & \textbf{Application}
    & \thead{\textbf{Arbitrary Number}\\\textbf{of Requests}} \\
    \midrule
    Su \etal~\cite{su2024towards}              & \CIRCLE & -- & -- & 100kB \\
    \rowcolor{white}
    Davis \etal~\cite{davis_case_2017}         & \CIRCLE & -- & -- & -- \\
    Turoňová \etal~\cite{turonova_counting_nodate}
                                               & \CIRCLE & -- & -- & 1.5K–9K \\
    \rowcolor{white}
    Liu \etal~\cite{liu_revealer_2021}         & \CIRCLE & -- & -- & 128 \\
    Staicu \etal~\cite{staicu_freezing_nodate} & --       & \CIRCLE & \CIRCLE & 82K \\
    \rowcolor{white}
    Davis \etal~\cite{davis_impact_2018}      & \CIRCLE & \CIRCLE & -- & 100K–1M \\
    Siddiq \etal~\cite{siddiq_understanding_2024}
                                               & \CIRCLE & -- & -- & -- \\
    \rowcolor{white}
    Shen \etal~\cite{shen_rescue_2018}         & \CIRCLE & -- & -- & 128 \\
    Barlas \etal~\cite{barlas_exploiting_2022} & --       & -- & \CIRCLE & -- \\
    \rowcolor{white}
    Siddiq \etal~\cite{siddiq_regexdoseval_2024}
                                               & \CIRCLE & -- & -- & -- \\
    Davis \etal~\cite{davis_why_2019}          & \CIRCLE & -- & -- & -- \\
    \rowcolor{white}
    Kluban \etal~\cite{kluban_measuring_2022}  & \CIRCLE & -- & -- & -- \\
    Bhuiyan \etal~\cite{bhuiyan_secbenchjs_2023}
                                               & \CIRCLE & \CIRCLE & -- & -- \\
    \rowcolor{white}
    Noller \etal~\cite{noller_badger_2018}     & \CIRCLE & -- & -- & -- \\
    Wang \etal~\cite{wang_effective_2023}      & \CIRCLE & \CIRCLE & -- & -- \\
    \rowcolor{white}
    Rathnayake \etal~\cite{rathnayake_static_2017}
                                               & \CIRCLE & -- & -- & -- \\
    McLaughlin \etal~\cite{mclaughlin_regulator_nodate}
                                               & \CIRCLE & \CIRCLE & -- & 1M \\
    \rowcolor{white}
    Li \etal~\cite{li_redoshunter_nodate}     & \CIRCLE & \CIRCLE & -- & -- \\
    Parolini \etal~\cite{parolini_sound_2023}  & \CIRCLE & -- & -- & 128 \\
    \rowcolor{white}
    Kirrage \etal~\cite{kirrage_static_2013}  & \CIRCLE & -- & -- & -- \\
    Wüstholz \etal~\cite{wustholz_static_2017}
                                               & \CIRCLE & -- & \CIRCLE & 140K \\
    \rowcolor{white}
    Petsios \etal~\cite{petsios_slowfuzz_2017} & \CIRCLE & -- & -- & -- \\
    Kluban \etal~\cite{kluban_detecting_2024}  & --       & \CIRCLE & -- & -- \\
    \rowcolor{white}
    Demoulin \etal~\cite{demoulin2019detecting}& --       & -- & \CIRCLE & -- \\
    Chida \etal~\cite{chida_repairing_2022}    & \CIRCLE & -- & -- & -- \\
    \rowcolor{white}
    Bai \etal~\cite{bai_runtime_2021}          & --       & -- & \CIRCLE & 50K \\
    Li \etal~\cite{li_regexscalpel_nodate}     & \CIRCLE & \CIRCLE & -- & 1M \\
    \rowcolor{white}
    Davis \etal~\cite{davis_rethinking_2019}   & \CIRCLE & -- & -- & -- \\
    Li \etal~\cite{li_flashregex_nodate}       & \CIRCLE & -- & -- & -- \\
    \rowcolor{white}
    Hassan \etal~\cite{hassan_improving_2023}  & \CIRCLE & -- & -- & -- \\
    Davis \etal~\cite{davis_sense_nodate}      & --       & \CIRCLE & \CIRCLE & -- \\
    \rowcolor{white}
    Davis \etal~\cite{davis_using_2021}        & \CIRCLE & -- & -- & 10K–100K \\
    Tandon \etal~\cite{tandon_leader_2023}     & --       & -- & \CIRCLE & -- \\
    \rowcolor{white}
    Weideman \etal~\cite{weideman_analyzing_2016}
                                               & \CIRCLE & -- & -- & -- \\
    Turoňová \etal~\cite{turonova_regex_2020}  & \CIRCLE & -- & -- & 500K–50M \\
    \bottomrule
  \end{tabular}%
}
\end{table}}


A realistic threat model is essential in any security study, as it provides a framework for assessing the potential findings. That is, it enables the identification of the system's weaknesses (\ie vulnerabilities) that adversaries could exploit in real-world environments. A strong threat model might lead to unrealistic findings, assuming too powerful attackers.
In ~Table~\ref{tab:threat_model}, we outline the granularity at which different threat models are defined in the analyzed papers. While most studies examine the exploitability of regexes in isolation, some do extend their analysis to the surrounding context, such as within library code. However, only a few studies define threat models at the application level. Threat models defined at the code fragment level can lead to false positives---vulnerabilities that are exploitable in isolated regexes but unreachable in actual code.

Among the papers we reviewed, only 10 (28\%) explicitly define a threat model, while the rest implicitly assume one. We identify four assumptions in the existing threat models:

    \myparagraph{Attacker Controls Input (\numberofpaper studies):} The assumption that an attacker can control inputs to a vulnerable ReDoS location is crucial for exploitability, as highlighted in various studies \cite{barlas_exploiting_2022, hassan_improving_2023}. In real-world software, including web applications and systems, inputs often originate from untrusted sources, thereby increasing the potential for exploitation \cite{petsios_slowfuzz_2017}. However, whether these inputs can reach the vulnerable regex depends on the surrounding code. All the reviewed papers, whether their threat models focus on the application~\cite{staicu_freezing_nodate, barlas_exploiting_2022, wustholz_static_2017, demoulin2019detecting}, library~\cite{bhuiyan_secbenchjs_2023, wang_effective_2023, mclaughlin_regulator_nodate, li_redoshunter_nodate}, or code fragment level~\cite{davis_case_2017, turonova_counting_nodate,liu_revealer_2021}, assume that the attacker can manipulate the input to exploit the vulnerability in the regex. However, this assumption may not always be applicable. For instance, consider a library that only reads and parses a configuration file on a server, which is not controlled by an attacker. In such cases, the likelihood of an attacker being able to influence the input and exploit the ReDoS vulnerability is significantly reduced, if not eliminated.

    \myparagraph{Attacker's Input Reaches Super-Linear Regex (34 studies):} This condition necessitates that the attacker's input be matched against a specific type of regex. While all the papers considering threat models at the library and code fragment levels implicitly assume this condition, it is a significant assumption. With the exception of Barlas \etal~\cite{barlas_exploiting_2022}, most studies presume this might be achievable in practice. In reality, achieving this could require the attacker to have access to detailed information about the server-side logic, API documentation, or the internal workings of the web service. Staicu \etal~\cite{staicu_freezing_nodate} incorporate web server implementations into their threat model and examine whether the input can trigger the vulnerability. However, this fingerprinting approach can be noisy, leading to both false positives and false negatives.

    \myparagraph{Use of Backtracking or Slow Regex Engine (33 studies):} This condition assumes the attacker has in-depth knowledge of the specific regex engine employed by the server. 
    On one hand, it might be feasible to remotely determine the type of engine in use (\eg PCRE, RE2), but pinpointing the exact version with its specific weaknesses can be quite challenging. 
    On the other hand, assuming a slow or outdated engine might lead to inaccurate results, making the attack scenario unrealistic in many real-world situations. Except for Barlas \etal~\cite{barlas_exploiting_2022} and Turo\v{n}ová \etal~\cite{turonova_counting_nodate}, all the studies make this assumption as part of their threat model.

    \myparagraph{Arbitrary Request Assumption (32 studies):} This condition represents a strong assumption that the attacker can repeatedly send requests with large input lengths without interference from other security mechanisms. Most studies, with the exceptions of \cite{shen_rescue_2018, liu_revealer_2021, parolini_sound_2023}, either disregard restrictions on request size and frequency or necessitate large inputs to trigger vulnerabilities. For context, the maximum size of HTTP header requests and responses in \texttt{nginx} and \texttt{Apache Tomcat}~\cite{apache_header, nginx_header} is 8K, which is significantly smaller than the input lengths assumed in most works. For instance, Davis \etal~\cite{davis_impact_2018} require input lengths of 100K-1M to trigger a 10-second slowdown, which would be impractical in real-world system settings. Furthermore, most studies fail to consider potential mitigation strategies employed by real-world systems, such as rate limiting, intrusion detection systems, or firewalls, which can significantly impede attacks. In realistic scenarios, attackers might face various system-imposed limitations, including bandwidth constraints or restrictions on the number of requests~\cite{heroku_header} they can send. As a result, the majority of studies, except for \cite{shen_rescue_2018, liu_revealer_2021, su2024towards}, assume in their threat models that attackers can perform requests unhindered, which may not reflect real-world conditions.

While the inclusion of a realistic threat model is essential for evaluating security findings accurately, many existing ReDoS studies make unrealistic, strong assumptions about attacker capabilities and system configurations.\enlargethispage{12pt}


\subsection{ReDoS Definitions}
\label{subsec:redos-definitions}


\aptLtoX{\begin{table}[t]
\normalsize
\centering
\caption{Summary of ReDoS vulnerability definitions in reviewed studies. The literature uses inconsistent criteria, with over half (19/29) defining ReDoS based on slowdown. \LEFTcircle~denotes incomplete details; \CIRCLE~indicates use of heuristics; -- means the study did not use or report that criterion.}
\label{tab:vulnerability_definition}
\resizebox{0.95\linewidth}{!}{%
  \scriptsize
  \begin{tabular}{lccc}
    \toprule
    \rowcolor{white}
      \textbf{Paper}
    & \textbf{Slowdown}
       & \begin{tabular}[c]{@{}l@{}}\textbf{Number of}\\\textbf{Instructions}\end{tabular}
    & \textbf{Heuristics} \\
    \midrule
    Su \etal~\cite{su2024towards}                      & --           & >10000       & --      \\
    Davis \etal~\cite{davis_case_2017}                 & --           & --           & \CIRCLE \\
    Turoňová \etal~\cite{turonova_counting_nodate}     & 10s, 100s    & --           & --      \\
    Liu \etal~\cite{liu_revealer_2021}                 & --           & $10^5$       & --      \\
    Staicu \etal~\cite{staicu_freezing_nodate}         & 5s           & --           & --      \\
    Davis \etal~\cite{davis_impact_2018}               & 10s          & --           & --      \\
    Siddiq \etal~\cite{siddiq_understanding_2024}      & 1s           & --           & --      \\
    Shen \etal~\cite{shen_rescue_2018}                 & --           & $10^8$       & --      \\
    Barlas \etal~\cite{barlas_exploiting_2022}         & 1s           & --           & --      \\
    Siddiq \etal~\cite{siddiq_regexdoseval_2024}       & 1s           & $10^8$       & --      \\
    Davis \etal~\cite{davis_why_2019}                  & 10s          & --           & --      \\
    Kluban \etal~\cite{kluban_measuring_2022}          & --           & --           & \CIRCLE \\
    Bhuiyan \etal~\cite{bhuiyan_secbenchjs_2023}       & 2s           & --           & --      \\
    Wang \etal~\cite{wang_effective_2023}              & --           & $10^5$       & --      \\
    McLaughlin \etal~\cite{mclaughlin_regulator_nodate}& 10s          & --           & --      \\
    Li \etal~\cite{li_redoshunter_nodate}              & 1s           & --           & --      \\
    Parolini \etal~\cite{parolini_sound_2023}          & --           & $10^{10}$    & --      \\
    Wüstholz \etal~\cite{wustholz_static_2017}         & \LEFTcircle  & --           & --      \\
    Petsios \etal~\cite{petsios_slowfuzz_2017}         & \LEFTcircle  & --           & --      \\
    Kluban \etal~\cite{kluban_detecting_2024}          & --           & --           & \CIRCLE \\
    Demoulin \etal~\cite{demoulin2019detecting}        & \LEFTcircle  & --           & --      \\
    Chida \etal~\cite{chida_repairing_2022}            & --           & --           & \CIRCLE \\
    Bai \etal~\cite{bai_runtime_2021}                  & $\mu + 3\sigma$ & --        & --      \\
    Li \etal~\cite{li_regexscalpel_nodate}             & 10s          & --           & --      \\
    Li \etal~\cite{li_flashregex_nodate}               & \LEFTcircle  & --           & --      \\
    Hassan \etal~\cite{hassan_improving_2023}          & --           & --           & \CIRCLE \\
    Davis \etal~\cite{davis_sense_nodate}              & 1s           & --           & --      \\
    Davis \etal~\cite{davis_using_2021}                & \LEFTcircle  & --           & --      \\
    Turoňová \etal~\cite{turonova_regex_2020}          & 10s, 100s    & --           & --      \\
    \bottomrule
  \end{tabular}%
}
\end{table}}{
\begin{table}[t]
\normalsize
\centering
\caption{Summary of ReDoS vulnerability definitions in reviewed studies. The literature uses inconsistent criteria, with over half (19/29) defining ReDoS based on slowdown. \LEFTcircle~denotes incomplete details; \CIRCLE~indicates use of heuristics; -- means the study did not use or report that criterion.}
\label{tab:vulnerability_definition}
\resizebox{0.95\linewidth}{!}{%
  \scriptsize
  \begin{tabular}{lGGG}
    \toprule
    \rowcolor{white}
      \textbf{Paper}
    & \textbf{Slowdown}
    & \thead{\textbf{Number of}\\\textbf{Instructions}}
    & \textbf{Heuristics} \\
    \midrule
    Su \etal~\cite{su2024towards}                      & --           & >10000       & --      \\
    \rowcolor{white}
    Davis \etal~\cite{davis_case_2017}                 & --           & --           & \CIRCLE \\
    Turoňová \etal~\cite{turonova_counting_nodate}     & 10s, 100s    & --           & --      \\
    \rowcolor{white}
    Liu \etal~\cite{liu_revealer_2021}                 & --           & $10^5$       & --      \\
    Staicu \etal~\cite{staicu_freezing_nodate}         & 5s           & --           & --      \\
    \rowcolor{white}
    Davis \etal~\cite{davis_impact_2018}               & 10s          & --           & --      \\
    Siddiq \etal~\cite{siddiq_understanding_2024}      & 1s           & --           & --      \\
    \rowcolor{white}
    Shen \etal~\cite{shen_rescue_2018}                 & --           & $10^8$       & --      \\
    Barlas \etal~\cite{barlas_exploiting_2022}         & 1s           & --           & --      \\
    \rowcolor{white}
    Siddiq \etal~\cite{siddiq_regexdoseval_2024}       & 1s           & $10^8$       & --      \\
    Davis \etal~\cite{davis_why_2019}                  & 10s          & --           & --      \\
    \rowcolor{white}
    Kluban \etal~\cite{kluban_measuring_2022}          & --           & --           & \CIRCLE \\
    Bhuiyan \etal~\cite{bhuiyan_secbenchjs_2023}       & 2s           & --           & --      \\
    \rowcolor{white}
    Wang \etal~\cite{wang_effective_2023}              & --           & $10^5$       & --      \\
    McLaughlin \etal~\cite{mclaughlin_regulator_nodate}& 10s          & --           & --      \\
    \rowcolor{white}
    Li \etal~\cite{li_redoshunter_nodate}              & 1s           & --           & --      \\
    Parolini \etal~\cite{parolini_sound_2023}          & --           & $10^{10}$    & --      \\
    \rowcolor{white}
    Wüstholz \etal~\cite{wustholz_static_2017}         & \LEFTcircle  & --           & --      \\
    Petsios \etal~\cite{petsios_slowfuzz_2017}         & \LEFTcircle  & --           & --      \\
    \rowcolor{white}
    Kluban \etal~\cite{kluban_detecting_2024}          & --           & --           & \CIRCLE \\
    Demoulin \etal~\cite{demoulin2019detecting}        & \LEFTcircle  & --           & --      \\
    \rowcolor{white}
    Chida \etal~\cite{chida_repairing_2022}            & --           & --           & \CIRCLE \\
    Bai \etal~\cite{bai_runtime_2021}                  & $\mu + 3\sigma$ & --        & --      \\
    \rowcolor{white}
    Li \etal~\cite{li_regexscalpel_nodate}             & 10s          & --           & --      \\
    Li \etal~\cite{li_flashregex_nodate}               & \LEFTcircle  & --           & --      \\
    \rowcolor{white}
    Hassan \etal~\cite{hassan_improving_2023}          & --           & --           & \CIRCLE \\
    Davis \etal~\cite{davis_sense_nodate}              & 1s           & --           & --      \\
    \rowcolor{white}
    Davis \etal~\cite{davis_using_2021}                & \LEFTcircle  & --           & --      \\
    Turoňová \etal~\cite{turonova_regex_2020}          & 10s, 100s    & --           & --      \\
    \bottomrule
  \end{tabular}%
}
\end{table}}


Most of the considered papers define a ReDoS vulnerability in terms of the slowdown that can be caused directly, with a limited number of characters. This variation is summarized in~Table~\ref{tab:vulnerability_definition}. Notably, there is no consensus on what constitutes a vulnerability. For instance, some studies establish thresholds for the number of allowed operations during matching that differ significantly. Liu \etal~\cite{liu_revealer_2021} and Wang \etal~\cite{wang_effective_2023} use a threshold of $10^5$ instructions, while Shen \etal~\cite{shen_rescue_2018} and Siddiq \etal~\cite{siddiq_regexdoseval_2024} consider $10^8$ instructions, and Parolini \etal~\cite{parolini_sound_2023} set it at $10^{10}$ instructions. Adding to the confusion, Sung \etal~\cite{caron_how_2022} define vulnerabilities at $10^6$ and $10^7$ instructions. It remains unclear why these values are deemed appropriate, as heavy CPU loads do not always indicate the presence of a vulnerability.

Moreover, the inconsistency extends to the measurement of introduced slowdowns. Various papers use different metrics to quantify slowdowns, further complicating the landscape and hindering comparability across studies. Turoňová \etal~\cite{turonova_counting_nodate} measure slowdowns in $10s$ and $100s$, Staicu \etal~\cite{staicu_freezing_nodate} in $5s$, Davis \etal~\cite{davis_impact_2018, davis_why_2019} in $10s$, Siddiq \etal~\cite{siddiq_understanding_2024}, Li \etal~\cite{li_redoshunter_nodate} and Barlas \etal~\cite{barlas_exploiting_2022} in $1s$, Bhuiyan \etal~\cite{bhuiyan_secbenchjs_2023} in $2s$. Meanwhile, McLaughlin \etal~\cite{mclaughlin_regulator_nodate} use $10s$, and Bai \etal~\cite{bai_runtime_2021} define a problematic slowdown as the average response time ($\mu$) plus three times the standard deviation ($3\sigma$), to account for significant deviations from the norm. These excessively high thresholds likely miss many potential vulnerabilities that could cause significant issues under real-world conditions. In our experiments, we saw that under one-second slowdowns can be weaponized against a target server, with modest attacker resources~\cite{bhuiyan2022talefrozencloudsquantifying}.
Other papers rely on heuristics to define a vulnerability. For example, Davis \etal~\cite{davis_case_2017} utilize safe-regex~\cite{noauthor_davisjamsafe-regex_nodate},
which uses static analysis rules to determine whether a regex is vulnerable. These tools are unsound and report both false positives and negatives.

The lack of consensus on vulnerability thresholds, slowdown metrics, and measurement methodologies highlights significant inconsistencies in ReDoS research, underscoring the need for standardized definitions and evaluation criteria.

\subsection{Evaluations}
\label{subsec-evaluation}


\aptLtoX{\begin{table}[t]
\normalsize
\centering
\caption{
Summary of evaluation methods used in the reviewed studies.
Most works (33/34) measure the success of a ReDoS attack from the perspective of the attacker, rather than from its impact on benign users (\ie Quality of Service (QoS) degradation).
\CIRCLE~indicates that the criterion or platform was explicitly considered; \LEFTcircle~denotes incomplete details; -- means the study did not use or report that criterion.
}
\label{tab:evaluation_languages_complete}
\resizebox{\linewidth}{!}{%
  \begin{tabular}{lcc|ccccccccc}
    \toprule
    \rowcolor{white}
      \textbf{Paper}
    & \textbf{Attack Simulation}
  &  \begin{tabular}[c]{@{}l@{}}\textbf{Measure QoS}\\\textbf{Degradation}\end{tabular}
    & {\textbf{JavaScript}}
    & {\textbf{Java}}
    & {\textbf{Python}}
    & {\textbf{Ruby}}
    & {\textbf{PHP}}
    & {\textbf{Go}}
    & {\textbf{Perl}}
    & {\textbf{Rust}}
    & {\textbf{C\#}} \\
    \midrule
    Su \etal~\cite{su2024towards}                 & --         & --         & \CIRCLE & \CIRCLE & --       & --       & \CIRCLE & \CIRCLE & \CIRCLE & \CIRCLE & \CIRCLE \\
    \rowcolor{white}
    Davis \etal~\cite{davis_case_2017}            & --         & --         & \CIRCLE & --      & --       & --       & --      & --      & --      & --      & --      \\
    Turoňová \etal~\cite{turonova_counting_nodate}& --         & --         & \CIRCLE & \CIRCLE & \CIRCLE & \CIRCLE & \CIRCLE & --      & \CIRCLE & --      & \CIRCLE \\
    \rowcolor{white}
    Liu \etal~\cite{liu_revealer_2021}            & --         & --         & \CIRCLE & \CIRCLE & \CIRCLE & \CIRCLE & \CIRCLE & --      & --      & --      & --      \\
    Staicu \etal~\cite{staicu_freezing_nodate}    & \LEFTcircle& --         & \CIRCLE & --      & --       & --       & --      & --      & --      & --      & --      \\
    \rowcolor{white}
    Davis \etal~\cite{davis_impact_2018}          & --         & --         & \CIRCLE & --      & \CIRCLE & --       & --      & --      & --      & --      & --      \\
    Siddiq \etal~\cite{siddiq_understanding_2024} & --         & --         & --      & \CIRCLE & --       & --       & --      & --      & --      & --      & --      \\
    \rowcolor{white}
    Shen \etal~\cite{shen_rescue_2018}            & --         & --         & --      & \CIRCLE & --       & --       & --      & --      & --      & --      & --      \\
    Barlas \etal~\cite{barlas_exploiting_2022}    & \LEFTcircle& --         & \CIRCLE & --      & --       & --       & --      & --      & --      & --      & --      \\
    \rowcolor{white}
    Siddiq \etal~\cite{siddiq_regexdoseval_2024}  & --         & --         & --      & \CIRCLE & --       & --       & --      & --      & --      & --      & --      \\
    Davis \etal~\cite{davis_why_2019}             & --         & --         & \CIRCLE & \CIRCLE & \CIRCLE & \CIRCLE & \CIRCLE & \CIRCLE & \CIRCLE & \CIRCLE & --      \\
    \rowcolor{white}
    Kluban \etal~\cite{kluban_measuring_2022}     & --         & --         & \CIRCLE & --      & --       & --       & --      & --      & --      & --      & --      \\
    Bhuiyan \etal~\cite{bhuiyan_secbenchjs_2023}  & --         & --         & \CIRCLE & --      & --       & --       & --      & --      & --      & --      & --      \\
    \rowcolor{white}
    Noller \etal~\cite{noller_badger_2018}        & --         & --         & --      & \CIRCLE & --       & --       & --      & --      & --      & --      & --      \\
    Wang \etal~\cite{wang_effective_2023}         & --         & --         & \CIRCLE & \CIRCLE & \CIRCLE & --       & --      & --      & --      & --      & \CIRCLE \\
    \rowcolor{white}
    Rathnayake \etal~\cite{rathnayake_static_2017}& --         & --         & --      & --      & \CIRCLE & --       & --      & --      & --      & --      & --      \\
    McLaughlin \etal~\cite{mclaughlin_regulator_nodate}
                                                  & --         & --         & \CIRCLE & --      & --       & --       & --      & --      & --      & --      & --      \\
    \rowcolor{white}
    Li \etal~\cite{li_redoshunter_nodate}         & --         & --         & --      & \CIRCLE & --       & --       & --      & --      & --      & --      & --      \\
    Parolini \etal~\cite{parolini_sound_2023}     & --         & --         & --      & \CIRCLE & --       & --       & --      & --      & --      & --      & --      \\
    \rowcolor{white}
    Kirrage \etal~\cite{kirrage_static_2013}     & --         & --         & --      & \CIRCLE & --       & --       & --      & --      & --      & --      & --      \\
    Wüstholz \etal~\cite{wustholz_static_2017}    & \LEFTcircle& --         & --      & \CIRCLE & --       & --       & --      & --      & --      & --      & --      \\
    \rowcolor{white}
    Petsios \etal~\cite{petsios_slowfuzz_2017}    & --         & --         & --      & --      & --       & \CIRCLE & --      & --      & --      & --      & --      \\
    Kluban \etal~\cite{kluban_detecting_2024}     & --         & --         & \CIRCLE & --      & --       & --       & --      & --      & --      & --      & --      \\
    \rowcolor{white}
    Demoulin \etal~\cite{demoulin2019detecting}   & \CIRCLE    & --         & \CIRCLE & --      & --       & --       & --      & --      & --      & --      & --      \\
    Chida \etal~\cite{chida_repairing_2022}       & --         & --         & \CIRCLE & --      & \CIRCLE & --       & --      & --      & --      & --      & --      \\
    \rowcolor{white}
    Bai \etal~\cite{bai_runtime_2021}             & \CIRCLE    & \CIRCLE    & \CIRCLE & --      & --       & --       & --      & --      & --      & --      & --      \\
    Li \etal~\cite{li_regexscalpel_nodate}        & --         & --         & --      & \CIRCLE & --       & --       & --      & --      & --      & --      & --      \\
    \rowcolor{white}
    Davis \etal~\cite{davis_rethinking_2019}      & --         & --         & \CIRCLE & \CIRCLE & \CIRCLE & \CIRCLE & \CIRCLE & \CIRCLE & \CIRCLE & \CIRCLE & --      \\
    Li \etal~\cite{li_flashregex_nodate}          & --         & --         & --      & \CIRCLE & --       & --       & --      & --      & --      & --      & --      \\
    \rowcolor{white}
    Hassan \etal~\cite{hassan_improving_2023}     & --         & --         & \CIRCLE & \CIRCLE & \CIRCLE & \CIRCLE & \CIRCLE & --      & \CIRCLE & --      & \CIRCLE \\
    Davis \etal~\cite{davis_sense_nodate}         & \LEFTcircle& --         & \CIRCLE & --      & --       & --       & --      & --      & --      & --      & --      \\
    \rowcolor{white}
    Tandon \etal~\cite{tandon_leader_2023}        & \CIRCLE    & --         & --      & --      & --       & --       & \CIRCLE & --      & --      & --      & --      \\
    Weideman \etal~\cite{weideman_analyzing_2016} & --         & --         & --      & \CIRCLE & --       & --       & --      & --      & --      & --      & --      \\
    \rowcolor{white}
    Turoňová \etal~\cite{turonova_regex_2020}     & --         & --         & --      & --      & --       & --       & --      & --      & --      & --      & \CIRCLE \\
    \bottomrule
  \end{tabular}%
}
\end{table}}{
\begin{table}[t]
\normalsize
\centering
\caption{
Summary of evaluation methods used in the reviewed studies.
Most works (33/34) measure the success of a ReDoS attack from the perspective of the attacker, rather than from its impact on benign users (\ie Quality of Service (QoS) degradation).
\CIRCLE~indicates that the criterion or platform was explicitly considered; \LEFTcircle~denotes incomplete details; -- means the study did not use or report that criterion.
}
\label{tab:evaluation_languages_complete}
\resizebox{\linewidth}{!}{%
  \begin{tabular}{lHH|HHHHHHHHH}
    \toprule
    \rowcolor{white}
      \textbf{Paper}
    & \textbf{Attack Simulation}
    & \thead{\textbf{Measure QoS}\\\textbf{Degradation}}
    & \rotatebox{90}{\textbf{JavaScript}}
    & \rotatebox{90}{\textbf{Java}}
    & \rotatebox{90}{\textbf{Python}}
    & \rotatebox{90}{\textbf{Ruby}}
    & \rotatebox{90}{\textbf{PHP}}
    & \rotatebox{90}{\textbf{Go}}
    & \rotatebox{90}{\textbf{Perl}}
    & \rotatebox{90}{\textbf{Rust}}
    & \rotatebox{90}{\textbf{C\#}} \\
    \midrule
    Su \etal~\cite{su2024towards}                 & --         & --         & \CIRCLE & \CIRCLE & --       & --       & \CIRCLE & \CIRCLE & \CIRCLE & \CIRCLE & \CIRCLE \\
    \rowcolor{white}
    Davis \etal~\cite{davis_case_2017}            & --         & --         & \CIRCLE & --      & --       & --       & --      & --      & --      & --      & --      \\
    Turoňová \etal~\cite{turonova_counting_nodate}& --         & --         & \CIRCLE & \CIRCLE & \CIRCLE & \CIRCLE & \CIRCLE & --      & \CIRCLE & --      & \CIRCLE \\
    \rowcolor{white}
    Liu \etal~\cite{liu_revealer_2021}            & --         & --         & \CIRCLE & \CIRCLE & \CIRCLE & \CIRCLE & \CIRCLE & --      & --      & --      & --      \\
    Staicu \etal~\cite{staicu_freezing_nodate}    & \LEFTcircle& --         & \CIRCLE & --      & --       & --       & --      & --      & --      & --      & --      \\
    \rowcolor{white}
    Davis \etal~\cite{davis_impact_2018}          & --         & --         & \CIRCLE & --      & \CIRCLE & --       & --      & --      & --      & --      & --      \\
    Siddiq \etal~\cite{siddiq_understanding_2024} & --         & --         & --      & \CIRCLE & --       & --       & --      & --      & --      & --      & --      \\
    \rowcolor{white}
    Shen \etal~\cite{shen_rescue_2018}            & --         & --         & --      & \CIRCLE & --       & --       & --      & --      & --      & --      & --      \\
    Barlas \etal~\cite{barlas_exploiting_2022}    & \LEFTcircle& --         & \CIRCLE & --      & --       & --       & --      & --      & --      & --      & --      \\
    \rowcolor{white}
    Siddiq \etal~\cite{siddiq_regexdoseval_2024}  & --         & --         & --      & \CIRCLE & --       & --       & --      & --      & --      & --      & --      \\
    Davis \etal~\cite{davis_why_2019}             & --         & --         & \CIRCLE & \CIRCLE & \CIRCLE & \CIRCLE & \CIRCLE & \CIRCLE & \CIRCLE & \CIRCLE & --      \\
    \rowcolor{white}
    Kluban \etal~\cite{kluban_measuring_2022}     & --         & --         & \CIRCLE & --      & --       & --       & --      & --      & --      & --      & --      \\
    Bhuiyan \etal~\cite{bhuiyan_secbenchjs_2023}  & --         & --         & \CIRCLE & --      & --       & --       & --      & --      & --      & --      & --      \\
    \rowcolor{white}
    Noller \etal~\cite{noller_badger_2018}        & --         & --         & --      & \CIRCLE & --       & --       & --      & --      & --      & --      & --      \\
    Wang \etal~\cite{wang_effective_2023}         & --         & --         & \CIRCLE & \CIRCLE & \CIRCLE & --       & --      & --      & --      & --      & \CIRCLE \\
    \rowcolor{white}
    Rathnayake \etal~\cite{rathnayake_static_2017}& --         & --         & --      & --      & \CIRCLE & --       & --      & --      & --      & --      & --      \\
    McLaughlin \etal~\cite{mclaughlin_regulator_nodate}
                                                  & --         & --         & \CIRCLE & --      & --       & --       & --      & --      & --      & --      & --      \\
    \rowcolor{white}
    Li \etal~\cite{li_redoshunter_nodate}         & --         & --         & --      & \CIRCLE & --       & --       & --      & --      & --      & --      & --      \\
    Parolini \etal~\cite{parolini_sound_2023}     & --         & --         & --      & \CIRCLE & --       & --       & --      & --      & --      & --      & --      \\
    \rowcolor{white}
    Kirrage \etal~\cite{kirrage_static_2013}     & --         & --         & --      & \CIRCLE & --       & --       & --      & --      & --      & --      & --      \\
    Wüstholz \etal~\cite{wustholz_static_2017}    & \LEFTcircle& --         & --      & \CIRCLE & --       & --       & --      & --      & --      & --      & --      \\
    \rowcolor{white}
    Petsios \etal~\cite{petsios_slowfuzz_2017}    & --         & --         & --      & --      & --       & \CIRCLE & --      & --      & --      & --      & --      \\
    Kluban \etal~\cite{kluban_detecting_2024}     & --         & --         & \CIRCLE & --      & --       & --       & --      & --      & --      & --      & --      \\
    \rowcolor{white}
    Demoulin \etal~\cite{demoulin2019detecting}   & \CIRCLE    & --         & \CIRCLE & --      & --       & --       & --      & --      & --      & --      & --      \\
    Chida \etal~\cite{chida_repairing_2022}       & --         & --         & \CIRCLE & --      & \CIRCLE & --       & --      & --      & --      & --      & --      \\
    \rowcolor{white}
    Bai \etal~\cite{bai_runtime_2021}             & \CIRCLE    & \CIRCLE    & \CIRCLE & --      & --       & --       & --      & --      & --      & --      & --      \\
    Li \etal~\cite{li_regexscalpel_nodate}        & --         & --         & --      & \CIRCLE & --       & --       & --      & --      & --      & --      & --      \\
    \rowcolor{white}
    Davis \etal~\cite{davis_rethinking_2019}      & --         & --         & \CIRCLE & \CIRCLE & \CIRCLE & \CIRCLE & \CIRCLE & \CIRCLE & \CIRCLE & \CIRCLE & --      \\
    Li \etal~\cite{li_flashregex_nodate}          & --         & --         & --      & \CIRCLE & --       & --       & --      & --      & --      & --      & --      \\
    \rowcolor{white}
    Hassan \etal~\cite{hassan_improving_2023}     & --         & --         & \CIRCLE & \CIRCLE & \CIRCLE & \CIRCLE & \CIRCLE & --      & \CIRCLE & --      & \CIRCLE \\
    Davis \etal~\cite{davis_sense_nodate}         & \LEFTcircle& --         & \CIRCLE & --      & --       & --       & --      & --      & --      & --      & --      \\
    \rowcolor{white}
    Tandon \etal~\cite{tandon_leader_2023}        & \CIRCLE    & --         & --      & --      & --       & --       & \CIRCLE & --      & --      & --      & --      \\
    Weideman \etal~\cite{weideman_analyzing_2016} & --         & --         & --      & \CIRCLE & --       & --       & --      & --      & --      & --      & --      \\
    \rowcolor{white}
    Turoňová \etal~\cite{turonova_regex_2020}     & --         & --         & --      & --      & --       & --       & --      & --      & --      & --      & \CIRCLE \\
    \bottomrule
  \end{tabular}%
}
\end{table}}


We carefully studied the evaluations of the papers in our dataset, as shown in Table~\ref{tab:evaluation_languages_complete}. Our analysis reveals two key aspects regarding ReDoS research:

\myparagraph{Language:} Regex matching is comparatively slow in programming languages like JavaScript, Java, Python, and Ruby~\cite{davis_why_2019}, leading to a significant amount of research focusing on detecting and defending ReDoS attacks in these languages. Reflecting this emphasis, 19 (54\%) papers address JavaScript, 18 (51\%) address Java, and 9 (25\%) address Python, with JavaScript being particularly prominent.

\myparagraph{Evaluation Methodology:} Out of the reviewed studies, only 7 (18\%), such as Demoulin\et~\cite{demoulin2019detecting}, Tandon\et~\cite{tandon_leader_2023}, and Staicu\et~\cite{staicu_freezing_nodate}, performed an end-to-end simulation to evaluate their proposed attack or defense methodologies. Moreover, only Bai\et~\cite{bai_runtime_2021} measured the system's performance degradation during an attack as perceived by benign users. The remaining 29 studies (82\%) rely solely on theoretical or localized analysis, which might not capture real-world attack scenarios and defenses accurately, potentially overestimating the prevalence of ReDoS vulnerabilities.

The lack of comprehensive end-to-end evaluations in current research hampers the ability to accurately gauge the real-world effectiveness and impact of ReDoS attacks and defenses.

\section{ReDoS Mitigations in Regex Engines}
\label{sec:engstudy}


In \S\ref{sec:litstudy}, we discuss many papers that measure ReDoS or propose defenses.
They often assume that the matching of a regex engine can be modeled as a backtracking search (\S\ref{subsec:threat-models}). However, it is unclear whether this assumption holds in practice, particularly in the context of modern regex engines.

In this section, we examine the regex engines of major programming languages to answer two questions:
  (1) \textit{Do the measurements and assumptions in the academic literature accurately reflect the latest versions of these engines?}
  and
  (2) \textit{What is the impact and adoption of ReDoS defenses proposed by researchers in real-world regex engines?}

We answer these questions in two ways.
First, a systematic analysis was conducted on regex engines in nine popular programming languages (\S\ref{sec:Engine-Analysis}). This analysis drew upon first-party sources such as source code, issue trackers, language documentation, blog posts from regex engine maintainers, and CVE reports. We also examined academic literature that describes regex engines. Second, we measured whether super-linear regexes improved performance in the latest versions of these engines (\S\ref{sec:Engines-Measurements}).

\subsection{Regex Engine Analysis} \label{sec:Engine-Analysis}
\BC{What to discuss here?} \JD{you can give some guidance about the structure of this part, eg ``we start by listing selection criteria, and then for each engine we note the X and the Y and look at the Z. The table summarizes.''} In this section, we begin by outlining the selection criteria used to choose the regex engines for our analysis. Then, for each engine, we examine the type and details of the ReDoS defense(s) implemented (if any), the update context of the defense (including the year of introduction), and the required lines of code (LOC) to implement it. Additionally, we assess whether the defense is enabled by default or requires developer intervention. We summarize our results on evaluating ReDoS defenses in modern regex engines in~Table~\ref{tab:EngineSummary}.




\aptLtoX{\begin{table*}[htbp]
\normalsize
\centering
\caption{
Summary of ReDoS defenses in the implementations of major programming languages.
}
\label{tab:EngineSummary}
\resizebox{\textwidth}{!}{%
  \begin{tabular}{ccccccc}
    \toprule
      \textbf{Language (\textit{Implementation})}
    & \textbf{Nature of the Defense}
    & \textbf{Deployment Year}
    & \textbf{LOC to Implement}
    & \textbf{On by Default?}
    & \textbf{Current ReDoS-Safeness} \\
    \midrule
    JavaScript (\textit{Node.js—V8} \cite{NodeJsNode, V8JavaScriptEngine})
      & A Non-backtracking Engine: Thompson-style
      & 2021
      & 1558 (10.6\%)
      & \xmark
      & Partial \\
    Ruby (\textit{MRI/CRuby} \cite{Ruby})
      & Memoization, Resource Capping: Time-based
      & 2022
      & 1100 (4.7\%)
      & \cmark
      & Partial \\
    C\# (\textit{.NET} \cite{WhatNETOpensource})
      & Resource Capping: Time-based; Brzozowski Derivatives Engine
      & 2012, 2022
      & 5417 (34.7\%)
      & \xmark
      & Partial \\
    Java (\textit{OpenJDK} \cite{OpenJDK})
      & Caching
      & 2016
      & 1712 (35.5\%)
      & \cmark
      & Partial \\
    PHP (\textit{Zend Engine} \cite{PHPGeneralInformation, phpPHPHistory})
      & Resource Capping: Counter-based
      & --
      & --
      & \cmark
      & Partial \\
    Perl (\textit{perl5} \cite{PerlWwwPerl})
      & Caching, Resource Capping: Counter-based
      & --
      & --
      & \cmark
      & Partial \\
    Rust (\textit{rustc} \cite{rustlangWhatRustc})
      & A Non-backtracking Engine: Thompson-style
      & --
      & --
      & \cmark
      & Safe \\
    Go (\textit{gc} \cite{IntroductionGoCompiler})
      & A Non-backtracking Engine: Thompson-style
      & --
      & --
      & \cmark
      & Safe \\
    Python (\textit{CPython} \cite{PythonDeveloperGuide})
      & --
      & --
      & --
      & \xmark
      & Not Safe \\
    \bottomrule
  \end{tabular}%
}
\end{table*}}{
\begin{table*}[htbp]
\normalsize
\centering
\caption{
Summary of ReDoS defenses in the implementations of major programming languages.
}
\label{tab:EngineSummary}
\resizebox{\textwidth}{!}{%
  \begin{tabular}{ccccccc}
    \toprule
    \rowcolor{white}
      \textbf{Language (\textit{Implementation})}
    & \textbf{Nature of the Defense}
    & \textbf{Deployment Year}
    & \textbf{LOC to Implement}
    & \textbf{On by Default?}
    & \textbf{Current ReDoS-Safeness} \\
    \midrule
    \rowcolor{table_gray!50}
    JavaScript (\textit{Node.js—V8} \cite{NodeJsNode, V8JavaScriptEngine})
      & A Non-backtracking Engine: Thompson-style
      & 2021
      & 1558 (10.6\%)
      & \xmark
      & Partial \\
    \rowcolor{white}
    Ruby (\textit{MRI/CRuby} \cite{Ruby})
      & Memoization, Resource Capping: Time-based
      & 2022
      & 1100 (4.7\%)
      & \cmark
      & Partial \\
    \rowcolor{table_gray!50}
    C\# (\textit{.NET} \cite{WhatNETOpensource})
      & Resource Capping: Time-based; Brzozowski Derivatives Engine
      & 2012, 2022
      & 5417 (34.7\%)
      & \xmark
      & Partial \\
    \rowcolor{white}
    Java (\textit{OpenJDK} \cite{OpenJDK})
      & Caching
      & 2016
      & 1712 (35.5\%)
      & \cmark
      & Partial \\
    \rowcolor{table_gray!50}
    PHP (\textit{Zend Engine} \cite{PHPGeneralInformation, phpPHPHistory})
      & Resource Capping: Counter-based
      & --
      & --
      & \cmark
      & Partial \\
    \rowcolor{white}
    Perl (\textit{perl5} \cite{PerlWwwPerl})
      & Caching, Resource Capping: Counter-based
      & --
      & --
      & \cmark
      & Partial \\
    \rowcolor{table_gray!50}
    Rust (\textit{rustc} \cite{rustlangWhatRustc})
      & A Non-backtracking Engine: Thompson-style
      & --
      & --
      & \cmark
      & Safe \\
    \rowcolor{white}
    Go (\textit{gc} \cite{IntroductionGoCompiler})
      & A Non-backtracking Engine: Thompson-style
      & --
      & --
      & \cmark
      & Safe \\
    \rowcolor{table_gray!50}
    Python (\textit{CPython} \cite{PythonDeveloperGuide})
      & --
      & --
      & --
      & \xmark
      & Not Safe \\
    \bottomrule
  \end{tabular}%
}
\end{table*}}


\myparagraph{Selection Criteria:}
Many modern programming languages have multiple implementations, runtime environments, engines, interpreters, or compilers that might use different regex engines. For instance, JavaScript has multiple engines for executing the source code, which can be used in different environments, such as V8 in Node.js and SpiderMonkey in Mozilla Firefox. To address this complexity, we applied two key criteria when selecting a specific regex engine for languages with multiple options. First, we prioritized engines with substantial real-world adoption, as these are the most likely targets of ReDoS attacks in practice. Second, in cases where usage data is sparse or fragmented, we selected reference implementations that are officially endorsed and openly maintained, enabling thorough analysis of ReDoS defenses. An overview of our regex engine selections for this study and their rationale can be found in Appendix~\ref{subsec:regex-engine-selection-summary}.

\subsubsection{JavaScript (\textit{Node.js---V8})}

\leavevmode\par\vspace{0.1\baselineskip}

\myparagraph{Original Algorithm:}
The main JavaScript engine, V8, previously used a backtracking Spencer-style algorithm.
It was highly optimized for common-case regexes~\cite{gruberSpeedingV8Regular2017}.
It uses an explicit automaton representation, optimizing it and then rendering it as native machine code for execution~\cite{corryIrregexpGoogleChrome}.

\myparagraph{ReDoS Defense:}
In 2021, responding to ReDoS concerns, the V8 developers implemented an extra Thompson-style non-backtracking regex engine~\cite{bidlingmaierAdditionalNonbacktrackingRegExp2021}.
The new engine guarantees linear-time complexity for all supported regexes, but as a trade-off, it does not support some E-regex features. All regexes with unsupported E-regex features still have to use the backtracking engine.
This update was introduced in V8 version 8.8, but in Node.js, it remains an experimental feature that is not enabled by default and must be activated via a feature flag.
Thus, some Node.js applications may not yet benefit from the new non-backtracking regex engine.

\myparagraph{Engineering Effort:}
The new engine is 1558 LOC. The original was 14741 LOC (excluding platform-specific assembly code). The effort was $\sim$10\% of the original.

\subsubsection{Ruby (\textit{MRI/CRuby})}

\leavevmode\par\vspace{0.1\baselineskip}

\myparagraph{Original Algorithm:}
Ruby's default regex engine is based on Onigmo \cite{k.takataKtakataOnigmo2024}, a fork of the Oniguruma \cite{k.kosakoKkosOniguruma2024} library, which is a Spencer-style backtracking regex engine.

\myparagraph{ReDoS Defense:}
Until Ruby 3.2 (2022), Ruby's port of Onigmo did not have built-in mitigations against ReDoS.
In Ruby 3.2, the developers added two ReDoS defenses~\cite{rubylangRuby320, ChangesRuby}.
First, based on a community proposal,\footnote{See \url{https://bugs.ruby-lang.org/issues/19104}.}
Ruby implemented Davis \etal's~\cite{davis_using_2021} memoization technique for Spencer's algorithm as a defense against ReDoS attacks.
Interestingly, the Ruby developers used full memoization rather than selective memoization despite the higher space complexity. Ruby does not apply memoization when unsupported E-regex features are encountered, instead offering a timeout mechanism similar to C\# as the second kind of defense.\footnote{See  \url{https://bugs.ruby-lang.org/issues/17837}.} The memoization defense is on by default.

\myparagraph{Engineering Effort:}
The relevant pull requests added a total of 1100 LOC, compared to the Ruby 3.1.6 engine's footprint of 23384 LOC. The ratio is 5\%.

\subsubsection{C\# (\textit{.NET)}}

\leavevmode\par\vspace{0.1\baselineskip}

\myparagraph{Original Algorithm:}
The first two regex engines in .NET used backtracking algorithms~\cite{BacktrackingNETRegular2023, RegularExpressionBehavior2021}. One was a simple opcode interpreter.
The other used an optimized machine code representation like V8's~\cite{BacktrackingNETRegular2023, toubPerformanceImprovementsNET2023, CompilationReuseRegular2021, gutierrezRegularExpressionPerformance2004}.

\myparagraph{ReDoS Defense:}
In .NET v4.5 (2012), .NET introduced a timeout mechanism to prevent excessive backtracking~\cite{WhatNewNET2023}. The timeout is checked after at most \( O(n) \) steps, where \( n \) is the length of the input, balancing fidelity with overhead~\cite{toubRegularExpressionImprovements2022, BestPracticesRegular2023}. In .NET 7 (2022), Microsoft added a non-backtracking regex engine~\cite{toubRegularExpressionImprovements2022}. This engine is based on Moseley \etal~ \cite{moseley_derivative_2023} and Saarikivi \etal~\cite{saarikiviSymbolicRegexMatcher2019}, with its core algorithm using Brzozowski derivatives and guaranteeing linear-time matches for supported regexes.
The new engine raises an error upon encountering unsupported E-regex features, so the developers must use the backtracking engine in such cases.
However, both the timeout mechanism and the non-backtracking engine are opt-in features.

\myparagraph{Engineering Effort:}
The new engine comprises 5417 LOC. The existing .NET codebase for regexes was 15597 LOC (including multiple engines). Hence, the new engine expands the existing regex engine codebase by 35\%. Microsoft researchers also created two research papers.

\subsubsection{Java (\textit{OpenJDK})}

\leavevmode\par\vspace{0.1\baselineskip}

\myparagraph{Original Algorithm:}
OpenJDK Java's previous regex engine implemented an NFA-based Spencer-style backtracking algorithm for regex matching~\cite{PatternJavaPlatform}.

\myparagraph{ReDoS Defense:}
As of Java 22 (2024), OpenJDK Java's regex engine does not document defenses against ReDoS. Despite the absence of official documentation, in the OpenJDK source code repository and issue tracker, we identified that some performance updates related to ReDoS issues were introduced in Java 9 (2016).\footnote{See \url{https://github.com/openjdk/jdk/commit/b45ea89}.} The OpenJDK maintainers introduced a bounded caching mechanism, improving problematic behavior for some ReDoS scenarios.
This ReDoS defense is enabled by default and affects all OpenJDK Java versions $\geq$9.

\myparagraph{Engineering Effort:}
Before the update, the regex-related codebase consisted of 4823 LOC. The memoization patch added 1712 lines (including empty lines and docstrings) to the Java 9 source code, which accounted for a 35\% expansion.

\subsubsection{Other ReDoS-Vulnerable Engines}

PHP (Zend Engine), Perl (perl5), and Python (CPython) are also known to be using Spencer-style backtracking regex engines \cite{friedl2006mastering, davis_why_2019}, which are vulnerable to ReDoS attacks.
According to Davis \etal~\cite{davis_using_2021}, PHP~\cite{PHPRuntimeConfiguration} and Perl~\cite{PerlMonks} utilize
``counter-based caps'' to prevent catastrophic backtracking as an alternative to C\# and Ruby's ``time-based caps''.
Python has neither built-in engine optimizations against ReDoS vulnerabilities nor a native timeout-like mechanism for regex evaluations.
We have not found any new documented updates or defenses against ReDoS for these languages since the analysis of Davis \etal~\cite{davis_using_2021} in 2021.

\subsubsection{Other ReDoS-Safe Engines}

Two major programming languages, Rust (rustc, via its official regex crate)~\cite{RustlangRegex2024} and Go (gc)~\cite{Goregex}, have regex engines that were developed with ReDoS safety in mind.
These engines simulate an explicit automaton representation using Thompson's algorithm~\cite{coxRegularExpressionMatching2007}.
They also do not support many E-regex features.
They, therefore, offer linear-time match guarantees in input and automaton length.

Although not included in our analysis, independent third-party regex engines like RE2 (Google)~\cite{GoogleRe22024} and Hyperscan (Intel)~\cite{wang_hyperscan_nodate, IntroductionHyperscan} are well-known for ReDoS resilience. RE2, a Thompson-style engine, incorporates DFA state caching optimizations~\cite{moseley_derivative_2023}. While generally ReDoS-safe and non-backtracking, RE2 can, in rare cases, exhibit backtracking behavior~\cite{davis_using_2021}. Similarly, Hyperscan is another efficient, ReDoS-safe engine, utilizing hardware acceleration alongside Glushkov's NFA construction~\cite{glushkovAbstractTheoryAutomata1961}. Neither engine supports E-regex features that require backtracking.

Overall, these developments in regex engines demonstrate significant progress in mitigating ReDoS risks but also highlight ongoing challenges in balancing the trade-off between safety and usability.
Modern regex engines mitigate ReDoS with non-backtracking algorithms like memoization, Thompson-style simulation, or Brzozowski derivatives to assure linear-time matching.
However, the benefit comes at the cost of reduced expressiveness.
Non-back\-track\-ing engines lack support for many E-regex features.
Defenses add timeouts or fallbacks for unsupported E-regexes to maintain codebase compatibility.\enlargethispage{12pt}

\subsection{Updated Measure of ReDoS Vulnerabilities} \label{sec:Engines-Measurements}

As indicated in~Table~\ref{tab:EngineSummary}, several major regex engines have recently been updated to mitigate or eliminate ReDoS.
We, therefore, wondered what proportion of ReDoS vulnerabilities have been addressed by state-of-the-art engines.

We answered this question through measurements on a sample from the Davis \etals~\cite{davis_why_2019} polyglot regex corpus, comprising $\sim$500K regexes extracted from open-source repositories across eight programming languages.
We used the tool vuln-regex-detector~\cite{davisDavisjamVulnregexdetector2024}, developed by Davis \etal~\cite{davis_impact_2018} and based on prior work~\cite{rathnayake_static_2017,wustholz_static_2017,weideman_analyzing_2016,shen_rescue_2018}, to identify potentially vulnerable regexes from the corpus.
From its predictions, we randomly sampled 500 regexes labeled as exponential-time candidates (out of $\sim$3K predicted examples) and 500 labeled as polynomial-time candidates (out of $\sim$100K predicted examples).
For each sampled regex, we used the same tool to generate corresponding attack input strings.

Then, we evaluated each regex-input pair on the regex engines described in~Table~\ref{tab:EngineSummary}, comparing the performance of the engine version before any changes to that of the most recent version.
If ReDoS defenses were available but not enabled by default in the latest version---as in the case of JavaScript and C\#---we activated them.
Following a parallel methodology that has been used in previous work~\cite{davis_why_2019,davis_impact_2018}, we define a regex's behavior as:
\begin{itemize}[leftmargin=10pt]
\item    \textit{Exponential}: The processing time exceeds five seconds with fewer than 500 pumps of an attack string.
\item    \textit{High-Polynomial}: The processing time exceeds five seconds with 500 or more pumps of an attack string.
\item    \textit{Low-Polynomial}: The processing time exceeds 0.5 seconds on an attack string but does not meet the exponential or high-polynomial behavior criteria.
\item    \textit{Linear}: The regex match never times out on the attack string.
\end{itemize}

\begin{figure}
    \centering
    \includegraphics[width=\linewidth]{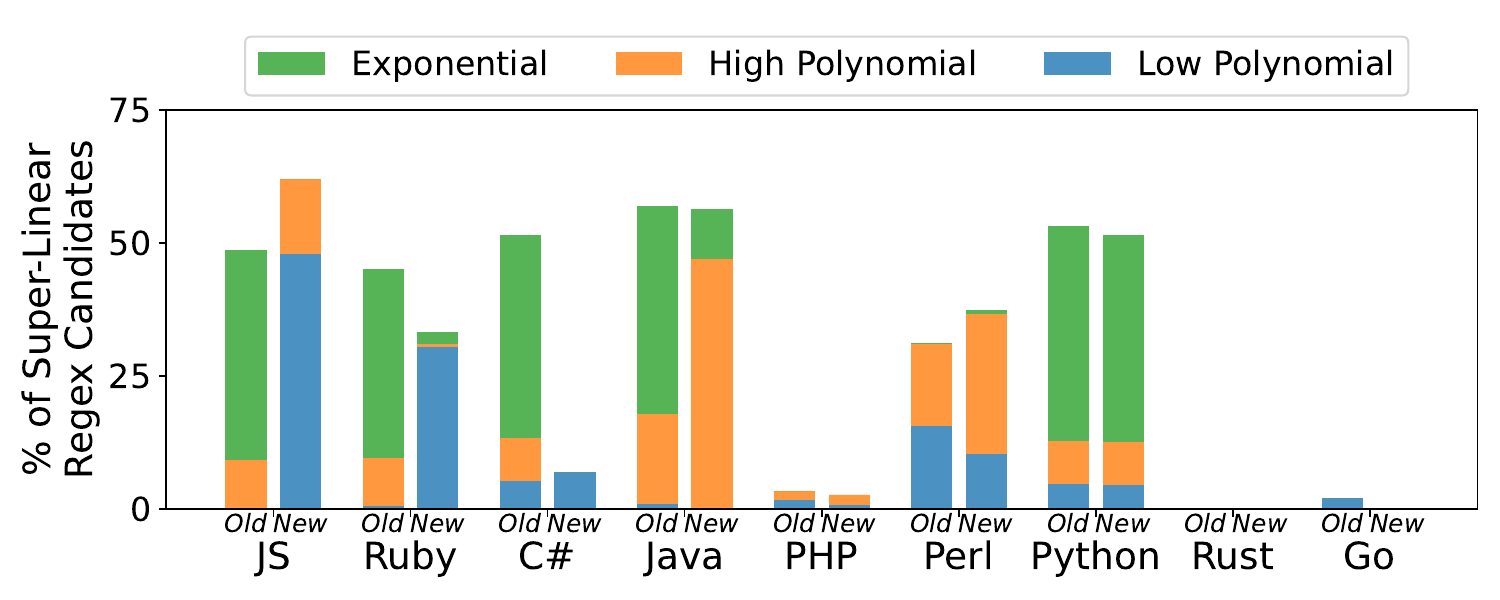}
    \caption{
    Performance of regex matching in old \vs latest engines in different programming languages. The exact engine versions are listed in Appendix~\ref{sec:engine-versions}.
    }
    \label{fig:UpdatedMeasureOfREDOS}

\end{figure}

Our findings are illustrated in Figure~\ref{fig:UpdatedMeasureOfREDOS}.
For each engine, we depict the number of regexes that exhibited super-linear behavior (as a fraction of the predicted input) in the pre- and post-ReDoS mitigation versions.
Not all regexes were supported in all engines, so the candidate pool size is indicated for each engine.
Observe that for the engines with mitigations, there is a substantial decrease in the number of viable super-linear regexes---in other words, these mitigations appear to be effective in addressing the common causes of super-linear behavior.
Notably, exponential behavior is resolved in C\# and JavaScript but persists in Ruby and Java, while polynomial behavior continues to manifest in JavaScript and Java.

\section{Discussion}
\label{sec:discussion}


In this section, we distill and contextualize the main findings of our study, integrating insights from our literature review (\S\ref{sec:litstudy}) and engineering review (\S\ref{sec:engstudy}).
For each idea, we discuss concrete future work directions for the community to explore.
\BC{TODO: Analyze combination of defenses in regex engines (see Slack)}
\BC{TODO: Propose a unified definition of ReDoS (see Slack)}

\myparagraph{(1) ReDoS is a peculiar vulnerability:}
Unlike traditional ones
like SQL injections, ReDoS
stands out as a contextual issue, more similar to micro-architectural attacks, where exploitability and impact are heavily dependent on the underlying system's architecture and deployment conditions.
For example, a regex pattern vulnerable to catastrophic backtracking in older versions of Node.js is rendered safe in newer versions due to advancements in regex engines (\S\ref{sec:engstudy}).
When detecting ReDoS vulnerabilities, it is not enough to analyze the code of a web application, without considering its deployment configurations.
For future static analysis work on ReDoS---particularly those targeting multiple languages such as ~\cite{hassan_improving_2023}---we recommend that they consider and model the language-, engine-, and platform-specific details and differences with respect to ReDoS.
For example, a super-linear regex might be more serious in a single-threaded web application architecture like Node.js than in inherently parallel ones like Go. 
For dynamic and hybrid analyses, we recommend a system-level perspective, similar to Shan\et~\cite{shan2017tail}, that deploys code under realistic conditions, taking relevant defenses into account.

\myparagraph{(2) Strong assumptions about attackers:}
%
We find that prior work often makes strong assumptions about attackers~(\S\ref{subsec:threat-models}), such as assuming they can inject arbitrarily large payloads or control the input to any regex.
Also, many papers neglect defenses or rely on rules of thumb, like regex matches should not exceed one second~(\S\ref{subsec:redos-definitions}).
We think that future work should challenge these assumptions by performing empirical studies with real systems, including analyses of how many regexes are actually reachable via attacker-controlled inputs.
We also need more empirical studies to show the community how modern applications are deployed in the wild---specifically, which configurations they run on and what defenses they employ.

\myparagraph{(3) Developers push back:}
\JD{We should be referencing the appendix in here?}
To the best of our knowledge, there has been no public description of a DoS attack that exploited ReDoS, so the risk posed by these vulnerabilities is still unclear, beyond academic measurement studies~\cite{staicu_freezing_nodate,barlas_exploiting_2022}.
Moreover, only a small fraction of the papers we analyzed~(\S\ref{sec:litstudy}) discuss such attacks, often in unrealistic setups.
We believe that these unclear risks together with problematic security assumptions are the culprits for ReDoS vulnerabilities being a contentious topic among developers.
Snyk has stated that ReDoS is the most neglected vulnerability type~\cite{openSourcereport},
implying that developers often ignore such findings. Also,
engineers at GitHub consider it a low severity vulnerability, ``{not particularly serious, but easy to create by accident, obscure to understand, and sometimes tricky to fix}''~\cite{githubblog}.
More strident voices argue that ReDoS vulnerabilities are ``{malicious noise}'' and just lead to security fatigue~\cite{trailofbitsguy}.
Several ReDoS CVE advisories have been disputed by the maintainers (\eg CVE-2023-39663, CVE-2022-42969, CVE-2019-11391), indicating disagreement between maintainers and reporters, also reflected by negative comments in GitHub issues~(Appendix~\ref{sec:comments-discussion}).

The blame for this perspective may lie with the cybersecurity academic community.
As noted in~\S\ref{sec:litstudy}, researchers have performed large-scale measurements of ReDoS as well as produced new and more effective detectors for vulnerable regexes.
Their measurement instruments have been integrated into popular security scanners (\eg the integration of safe-regex into eslint).
However, the resulting scanners analyze only the regex, without considering reachability constraints as well as other input constraints.
When academics apply these tools, they are likely to filter false positives based on the regex’s context.
In contrast, practitioners sometimes make much ado about nothing.
This suggests the need for empirical studies of academic tool uptake and more developer-friendly releases.
More importantly, future work in the field should perform user studies with developers to better understand their challenges related to ReDoS---such as tool usability, security assumptions, and deployment configurations.
In this way, we can better align the community's assumptions with those of practitioners.

\myparagraph{(4) Engines have evolved significantly:}
It is encouraging that modern language runtimes changed their implementation in response to prior work on ReDoS~(\S\ref{sec:engstudy}). These changes were often driven or informed by researchers (\eg Ruby's adoption of Davis \etals defenses~\cite{davis_rethinking_2019,davis_using_2021}).
We believe that this is a success story of effective collaboration between researchers and practitioners.

Our study shows that modern regex engines have evolved significantly, incorporating advanced techniques such as Thompson's algorithm, memoization, and resource capping to mitigate ReDoS.
However, there are often redundant implementations that are triggered on demand by the engine, when matching a regex.
Hence, future work should investigate the deployed fixes to find potential bugs or misbehaving corner cases.
We advocate for comparative studies (such as ours in \S\ref{sec:Engines-Measurements}), where multiple regex engines are evaluated on the same regexes or datasets to systematically assess the strengths and weaknesses of their deployed defenses.
The research community should also propose automatic solutions for migrating to updated engines.
At the same time, research on ReDoS should take these advancements into account to ensure its findings remain applicable.

\myparagraph{(5) Performance bugs vs. security vulnerabilities:}
Many of our findings may also extend to other algorithmic complexity vulnerabilities.
While the threat of weaponizing slow computations is real, in practice, attackers still tend to favor brute-force, volume-based distributed DoS attacks.
Researchers should aim to understand why that is the case and, simultaneously, continue to develop defenses against such potential future attacks. We also believe that there are many insufficiently explored, cross-disciplinary opportunities for ideas that aim to bridge the gap between performance, measurement, and security communities.
For example, we argue that most performance bugs (\eg redundant loop traversals~\cite{OlivoDL15}) can probably be lifted to a DoS attack when malicious intent is assumed.
Thus, future work should further examine the relation between performance bugs and DoS attacks.


\vspace{-2mm}
\section{Conclusion}


In this work, we perform a multi-faceted study of the recent academic work on ReDoS and the new engineering realities in the mainstream programming languages, after initial fixes were deployed in most of them. We report on concrete examples of language runtimes that adopted defenses, which appear to be motivated by academic research. When studying the efficacy of these defenses, we find that they mitigate most ReDoS payloads, but they still leave significant room for attackers to maneuver.
By surveying the academic work on ReDoS, we find that many papers in this domain use weak definitions and strong threat models, often considering any slowdown caused by regex matching as a security vulnerability. 
We advocate for future work to consider the recently-deployed fixes in mainstream programming languages and to evaluate ReDoS attacks under realistic deployment conditions (\eg against a web application deployed in the cloud using reversed proxies and redundancy). We also propose restricting the attackers' capabilities to realistic payload sizes and modest bandwidths that comply with the asymmetric DoS setting. More importantly, future work should study the more general problem of weaponizing slow computation, which the attackers might leverage in the near future.


\vspace{-0.03cm}
\begin{acks}
This work was supported in part by the \grantsponsor{YRXVL4JYCEF5}{National Science Foundation (NSF)}{} under grant
\#\grantnum{YRXVL4JYCEF5}{2135156}.
\end{acks}

\vspace{-0.03cm}
\section*{Ethical Considerations}

In our judgment, this work presents an acceptable ethical risk-reward tradeoff. Specifically, the literature review (\S\ref{sec:litstudy}) and engineering review (\S\ref{sec:engstudy}) describe and systematize only publicly available resources.

\vspace{-0.03cm}
\section*{Data Availability}

An artifact accompanying this paper is available at \url{https://doi.org/10.5281/zenodo.15515222}.
It includes: the CVE data analyzed in \S\ref{subsec:redos-in-os}, the super-linear regex corpus together with the reproduction scripts for the experiments described in \S\ref{sec:Engines-Measurements}, and the developer-discussion data and GitHub-crawler scripts from Appendix~\ref{sec:comments-discussion}.

\BC{OLD BUT PROBABLY STILL VALID: I put this comment here for visibility: The citation format for [74] is wrong, and [60] has changed its name to Repairing DoS Vulnerability of Real-World Regexes (https://arxiv.org/abs/2010.12450v4), so it needs an update in the bibliography.}

\JD{The bibliography has some issues that could fail a desk reject, notably URLs overflowing the columns}
\JD{Some duplicate bib entries: Cox's blog occurs twice.}
\JD{Missing URL -- [62] cheat sheet}
\JD{[70] is missing venue (and repeats with [71])}
\JD{[72] is missing venue}
\JD{Does [79] have a peer reviewed version?}
\JD{[82] has a peer reviewed version, use that}
\JD{[88] has a peer reviewed version, use that}
\JD{[89] and [90]: missing venue info}
\JD{[95] is PLDI I think}
\JD{[100]: missing venue info}
\JD{is there a peer reviewed version of [104]? maybe not}
\JD{[108]: missing venue info}
\JD{Missing URL -- [110]}
\JD{missing venue: 111}
\JD{missing venue: 112 (which USENIX conference?)}
\JD{diacritic mangling in [120]}
\JD{what is the source of 122?}
\JD{missing venue: 123}
\JD{diacritic mangling in 126}



\bibliographystyle{ACM-Reference-Format}




\appendix



\section*{Outline of the Appendices}

\noindent
The appendices contain the following material:

\begin{itemize}
\item Reference table of regex features and notation (Appendix~\ref{sec:appendix-regexSyntaxSemantics}).
\item Methodology on how we studied ReDoS in practice (Appendix~\ref{sec:appendix-REDOSInPractice}).
\item Information regarding the exclusion of works from the literature review (Appendix~\ref{subsec:excluded-works}).
\item Additional details of the engineering review, including engine selection and version information (Appendix~\ref{sec:appendix-engine-review}).
\item A look at the developer discussions about ReDoS vulnerabilities (Appendix~\ref{sec:comments-discussion}).
\end{itemize}

\section{Regex Feature Reference} \label{sec:appendix-regexSyntaxSemantics}

Table~\ref{table:Compact-PCRE-Notation} provides a comprehensive reference of regex features and notation, summarized from the PCRE2 specification~\cite{Hazel1997PCRE}. The table categorizes regex features into two groups: K-regexes and E-regexes. E-regexes are also subcategorized. Each feature is associated with its syntactic notation.

\renewcommand{\arraystretch}{1}

\aptLtoX{\begin{table}
    \footnotesize
    \centering
    \caption[Full set of features and notation of regexes]{%
        Full set of features and notation of regexes, compiled from~\cite{Hazel1997PCRE}. $R$ denotes a sub-pattern.}
    \begin{tabular}{ccc}
        \toprule
        \textbf{Abbreviation} & \textbf{Feature} & \textbf{Notation} \\ \midrule
        \multicolumn{3}{c}{\textbf{K-regexes}} \\ \midrule
         \multicolumn{1}{c|}{\cellcolor{table_gray!50}CAT}  & \multicolumn{1}{c|}{X followed by Y}      & $R_1 R_2$ \\
        \multicolumn{1}{c|}{KLE}  & \multicolumn{1}{c|}{Zero-or-more repetition} & $R*$ \\
          \multicolumn{1}{c|}{\cellcolor{table_gray!50}OR}   & \multicolumn{1}{c|}{Logical OR}           & $R_1|R_2$ \\ \midrule
        \multicolumn{3}{c}{\textbf{E-regexes}} \\ \midrule
        \multicolumn{3}{c}{\textbf{Capture Groups}} \\ \midrule
         \multicolumn{1}{c|}{\cellcolor{table_gray!50}CG}   & \multicolumn{1}{c|}{Capture group}          & $(R)$ \\
        \multicolumn{1}{c|}{NCG}  & \multicolumn{1}{c|}{Non-capture group}    & $(?:R)$ \\
         \multicolumn{1}{c|}{\cellcolor{table_gray!50}PNG} & \multicolumn{1}{c|}{Named capture group}    & $(?{\textless}\text{name}{\textgreater}R)$ \\ \midrule
        \multicolumn{3}{c}{\textbf{Quantifiers}} \\ \midrule
       \multicolumn{1}{c|}{\cellcolor{table_gray!50}ADD}  & \multicolumn{1}{c|}{One-or-more repetition} & $R+$ \\
        \multicolumn{1}{c|}{QST}  & \multicolumn{1}{c|}{Zero-or-one repetition} & $R?$ \\
        \multicolumn{1}{c|}{\cellcolor{table_gray!50}DBB}  & \multicolumn{1}{c|}{Range repetition}      & $R\{m,n\}$ \\
        \multicolumn{1}{c|}{LWB}  & \multicolumn{1}{c|}{At-least-$m$ repetition} & $R\{m,\}$ \\
         \multicolumn{1}{c|}{\cellcolor{table_gray!50}SNG}  & \multicolumn{1}{c|}{Exactly-$n$ repetition} & $R\{n\}$ \\ \midrule
        \multicolumn{3}{c}{\textbf{Character Classes}} \\ \midrule
         \multicolumn{1}{c|}{\cellcolor{table_gray!50}CCC}  & \multicolumn{1}{c|}{Custom character class} & $[\text{aeiou}]$ \\
        \multicolumn{1}{c|}{RNG}  & \multicolumn{1}{c|}{Character range}        & $[\text{a}-\text{z}]$ \\
        \multicolumn{1}{c|}{\cellcolor{table_gray!50}NCCC} & \multicolumn{1}{c|}{Negated class}         & \begin{imageonly}[\^{}$\text{aeiou}$]\end{imageonly} \\
        \multicolumn{1}{c|}{ANY}  & \multicolumn{1}{c|}{Any character}          & $.$ \\
        \multicolumn{1}{c|}{\cellcolor{table_gray!50}WSP}  & \multicolumn{1}{c|}{Whitespace}            & $\setminus \text{s}$ \\
        \multicolumn{1}{c|}{DEC}  & \multicolumn{1}{c|}{Numeric}                & $\setminus \text{d}$ \\
        \multicolumn{1}{c|}{\cellcolor{table_gray!50}WRD}  & \multicolumn{1}{c|}{Word}                 & $\setminus \text{w}$ \\
        \multicolumn{1}{c|}{NWSP} & \multicolumn{1}{c|}{Non-whitespace}         & $\setminus \text{S}$ \\
       \multicolumn{1}{c|}{\cellcolor{table_gray!50}NDEC} & \multicolumn{1}{c|}{Non-numeric}          & $\setminus \text{D}$ \\
        \multicolumn{1}{c|}{NWRD} & \multicolumn{1}{c|}{Non-word}               & $\setminus \text{W}$ \\
        \multicolumn{1}{c|}{\cellcolor{table_gray!50}VWSP} & \multicolumn{1}{c|}{Vertical space}       & $\setminus \text{v}$ \\ \midrule
        \multicolumn{3}{c}{\textbf{Zero-width Assertions}} \\ \midrule
        \multicolumn{1}{c|}{\cellcolor{table_gray!50}STR}  & \multicolumn{1}{c|}{Start-of-string/line} & \begin{imageonly}\^{}$R,\;\setminus\text{A}R$\end{imageonly} \\
        \multicolumn{1}{c|}{END}  & \multicolumn{1}{c|}{End-of-string/line}     & $R\$,\;R\setminus\text{Z}$ \\
        \multicolumn{1}{c|}{\cellcolor{table_gray!50}WNW}  & \multicolumn{1}{c|}{Word/non-word boundary} & $\setminus\text{b}$ \\
        \multicolumn{1}{c|}{NWNW} & \multicolumn{1}{c|}{Negated WNW boundary}   & $\setminus\text{B}$ \\
       \multicolumn{1}{c|}{\cellcolor{table_gray!50}PLA}  & \multicolumn{1}{c|}{Positive lookahead}  & $(?=R)$ \\
        \multicolumn{1}{c|}{NLA}  & \multicolumn{1}{c|}{Negative look-ahead}    & $(?!R)$ \\
        \multicolumn{1}{c|}{\cellcolor{table_gray!50}PLB}  & \multicolumn{1}{c|}{Positive lookbehind} & $(?{\textless}=R)$ \\
        \multicolumn{1}{c|}{NLB}  & \multicolumn{1}{c|}{Negative look-behind}   & $(?{\textless}!R)$ \\ \midrule
        \multicolumn{3}{c}{\textbf{Backreferences}} \\ \midrule
         \multicolumn{1}{c|}{\cellcolor{table_gray!50}BKR}  & \multicolumn{1}{c|}{Numeric backreference} & $(R)\ldots\setminus1$ \\
        \multicolumn{1}{c|}{BKRN} & \multicolumn{1}{c|}{Named backreference}     & $(?{\textless}\text{name}{\textgreater}R)\ldots\setminus\text{k{\textless}name{\textgreater}}$ \\ \midrule
        \multicolumn{3}{c}{\textbf{Backtracking Controls}} \\ \midrule
        \multicolumn{1}{c|}{\cellcolor{table_gray!50}LZY}  & \multicolumn{1}{c|}{Non-greedy repetition} & $R*?,\;R+?,\;R\{m,n\}?$ \\
        \multicolumn{1}{c|}{ATM}  & \multicolumn{1}{c|}{Atomic group}            & $(?{\textgreater}R)$ \\
        \multicolumn{1}{c|}{\cellcolor{table_gray!50}POS}  & \multicolumn{1}{c|}{Possessive quantifier} & $R++,\;R*+$ \\ \bottomrule
    \end{tabular}
    \label{table:Compact-PCRE-Notation}
\end{table}}{\begin{table}
    \footnotesize
    \centering
    \caption[Full set of features and notation of regexes]{%
        Full set of features and notation of regexes, compiled from~\cite{Hazel1997PCRE}. $R$ denotes a sub-pattern.}
    \begin{tabular}{ccc}
        \toprule
        \rowcolor{white}\textbf{Abbreviation} & \textbf{Feature} & \textbf{Notation} \\ \midrule
        \multicolumn{3}{c}{\textbf{K-regexes}} \\ \midrule
        \rowcolor{table_gray!50}\multicolumn{1}{c|}{CAT}  & \multicolumn{1}{c|}{X followed by Y}      & $R_1 R_2$ \\
        \multicolumn{1}{c|}{KLE}  & \multicolumn{1}{c|}{Zero-or-more repetition} & $R*$ \\
        \rowcolor{table_gray!50}\multicolumn{1}{c|}{OR}   & \multicolumn{1}{c|}{Logical OR}           & $R_1|R_2$ \\ \midrule
        \multicolumn{3}{c}{\textbf{E-regexes}} \\ \midrule
        \multicolumn{3}{c}{\textbf{Capture Groups}} \\ \midrule
        \rowcolor{table_gray!50}\multicolumn{1}{c|}{CG}   & \multicolumn{1}{c|}{Capture group}          & $(R)$ \\
        \multicolumn{1}{c|}{NCG}  & \multicolumn{1}{c|}{Non-capture group}    & $(?:R)$ \\
        \rowcolor{table_gray!50}\multicolumn{1}{c|}{PNG}  & \multicolumn{1}{c|}{Named capture group}    & $(?{\textless}\text{name}{\textgreater}R)$ \\ \midrule
        \multicolumn{3}{c}{\textbf{Quantifiers}} \\ \midrule
        \rowcolor{table_gray!50}\multicolumn{1}{c|}{ADD}  & \multicolumn{1}{c|}{One-or-more repetition} & $R+$ \\
        \multicolumn{1}{c|}{QST}  & \multicolumn{1}{c|}{Zero-or-one repetition} & $R?$ \\
        \rowcolor{table_gray!50}\multicolumn{1}{c|}{DBB}  & \multicolumn{1}{c|}{Range repetition}      & $R\{m,n\}$ \\
        \multicolumn{1}{c|}{LWB}  & \multicolumn{1}{c|}{At-least-$m$ repetition} & $R\{m,\}$ \\
        \rowcolor{table_gray!50}\multicolumn{1}{c|}{SNG}  & \multicolumn{1}{c|}{Exactly-$n$ repetition} & $R\{n\}$ \\ \midrule
        \multicolumn{3}{c}{\textbf{Character Classes}} \\ \midrule
        \rowcolor{table_gray!50}\multicolumn{1}{c|}{CCC}  & \multicolumn{1}{c|}{Custom character class} & $[\text{aeiou}]$ \\
        \multicolumn{1}{c|}{RNG}  & \multicolumn{1}{c|}{Character range}        & $[\text{a}-\text{z}]$ \\
        \rowcolor{table_gray!50}\multicolumn{1}{c|}{NCCC} & \multicolumn{1}{c|}{Negated class}         & $[\textasciicircum \text{aeiou}]$ \\
        \multicolumn{1}{c|}{ANY}  & \multicolumn{1}{c|}{Any character}          & $.$ \\
        \rowcolor{table_gray!50}\multicolumn{1}{c|}{WSP}  & \multicolumn{1}{c|}{Whitespace}            & $\setminus \text{s}$ \\
        \multicolumn{1}{c|}{DEC}  & \multicolumn{1}{c|}{Numeric}                & $\setminus \text{d}$ \\
        \rowcolor{table_gray!50}\multicolumn{1}{c|}{WRD}  & \multicolumn{1}{c|}{Word}                 & $\setminus \text{w}$ \\
        \multicolumn{1}{c|}{NWSP} & \multicolumn{1}{c|}{Non-whitespace}         & $\setminus \text{S}$ \\
        \rowcolor{table_gray!50}\multicolumn{1}{c|}{NDEC} & \multicolumn{1}{c|}{Non-numeric}          & $\setminus \text{D}$ \\
        \multicolumn{1}{c|}{NWRD} & \multicolumn{1}{c|}{Non-word}               & $\setminus \text{W}$ \\
        \rowcolor{table_gray!50}\multicolumn{1}{c|}{VWSP} & \multicolumn{1}{c|}{Vertical space}       & $\setminus \text{v}$ \\ \midrule
        \multicolumn{3}{c}{\textbf{Zero-width Assertions}} \\ \midrule
        \rowcolor{table_gray!50}\multicolumn{1}{c|}{STR}  & \multicolumn{1}{c|}{Start-of-string/line} & $\textasciicircum R,\;\setminus\text{A}R$ \\
        \multicolumn{1}{c|}{END}  & \multicolumn{1}{c|}{End-of-string/line}     & $R\$,\;R\setminus\text{Z}$ \\
        \rowcolor{table_gray!50}\multicolumn{1}{c|}{WNW}  & \multicolumn{1}{c|}{Word/non-word boundary} & $\setminus\text{b}$ \\
        \multicolumn{1}{c|}{NWNW} & \multicolumn{1}{c|}{Negated WNW boundary}   & $\setminus\text{B}$ \\
        \rowcolor{table_gray!50}\multicolumn{1}{c|}{PLA}  & \multicolumn{1}{c|}{Positive lookahead}  & $(?=R)$ \\
        \multicolumn{1}{c|}{NLA}  & \multicolumn{1}{c|}{Negative look-ahead}    & $(?!R)$ \\
        \rowcolor{table_gray!50}\multicolumn{1}{c|}{PLB}  & \multicolumn{1}{c|}{Positive lookbehind} & $(?{\textless}=R)$ \\
        \multicolumn{1}{c|}{NLB}  & \multicolumn{1}{c|}{Negative look-behind}   & $(?{\textless}!R)$ \\ \midrule
        \multicolumn{3}{c}{\textbf{Backreferences}} \\ \midrule
        \rowcolor{table_gray!50}\multicolumn{1}{c|}{BKR}  & \multicolumn{1}{c|}{Numeric backreference} & $(R)\ldots\setminus1$ \\
        \multicolumn{1}{c|}{BKRN} & \multicolumn{1}{c|}{Named backreference}     & $(?{\textless}\text{name}{\textgreater}R)\ldots\setminus\text{k{\textless}name{\textgreater}}$ \\ \midrule
        \multicolumn{3}{c}{\textbf{Backtracking Controls}} \\ \midrule
        \rowcolor{table_gray!50}\multicolumn{1}{c|}{LZY}  & \multicolumn{1}{c|}{Non-greedy repetition} & $R*?,\;R+?,\;R\{m,n\}?$ \\
        \multicolumn{1}{c|}{ATM}  & \multicolumn{1}{c|}{Atomic group}            & $(?{\textgreater}R)$ \\
        \rowcolor{table_gray!50}\multicolumn{1}{c|}{POS}  & \multicolumn{1}{c|}{Possessive quantifier} & $R++,\;R*+$ \\ \bottomrule
    \end{tabular}
    \label{table:Compact-PCRE-Notation}\vspace{-16pt}
\end{table}}

\begin{table}[!ht]
  \centering
  \caption{Examples of excluded works with explanations.}
  \label{tab:excluded_papers}
  \resizebox{\linewidth}{!}{%
    \begin{tabular}{p{3.2cm}p{6.8cm}}
      \toprule
      \textbf{Work} & \textbf{Reason for Exclusion} \\
      \midrule
      Cox~\cite{coxRegularExpressionMatching2007} &
        \cellcolor{table_gray!50}Evaluation does not address ReDoS-specific scenarios. \\
      Varatalu~\etal~\cite{varatalu2025re} &
        Evaluation does not address ReDoS-specific scenarios. \\
      Chattopadhyay~\etal~\cite{chattopadhyay2025verified} &
        \cellcolor{table_gray!50}Evaluation does not address ReDoS-specific scenarios. \\
      Mamouras~\etal~\cite{mamouras_efficient_2024} &
        Evaluation does not address ReDoS-specific scenarios. \\
      Shan~\etal~\cite{shan2017tail} &
        \cellcolor{table_gray!50}Addresses general DoS attacks. \\
      Wei~\etal~\cite{meng2018rampart} &
        Addresses general DoS attacks. \\
      \bottomrule
    \end{tabular}%
  }
\vspace{-20pt}
\end{table}

\begin{table*}
\centering
\caption{Regex engine implementations selected for evaluating ReDoS defenses.}
\label{tab:appendix-engine_selection}
\resizebox{\textwidth}{!}{
\begin{tabular}{cccc}
\toprule
\rowcolor{white}
\textbf{PL} & \textbf{Selected Impl. of the PL [Source Code]} & \textbf{Other Alternatives} & \textbf{Reason for Selection} \\
\midrule
 \rowcolor{table_gray!50}
JavaScript                         & Node.js—V8 \cite{NodejsNode2024, V8V8Git}            & Deno, Bun, SpiderMonkey, JavaScriptCore & Dominates server-side JavaScript usage (98.8\% market share) \cite{DistributionWebServersa}               \\
\rowcolor{white}
Ruby                               & MRI/CRuby \cite{RubyRuby2024}                         & JRuby, Rubinius, mruby, RubyMotion & Reference implementation \cite{Ruby}                \\
\rowcolor{table_gray!50}
C\#                                 & .NET \cite{DotnetRuntime2024}                         & Mono & Reference implementation \cite{LanguageSpecification2024}                \\
\rowcolor{white}
Java                               & OpenJDK \cite{OpenjdkJdk2024}                        & Amazon Corretto, Azul Zulu, Liberica JDK, Eclipse Adoptium... & Most widely used JDK of 2024 with a 20.8\% usage rate \cite{2024StateJava}                \\
\rowcolor{table_gray!50}
PHP                                & Zend Engine \cite{PhpPhpsrc2024a}                    & HHVM, PeachPie, Quercus, Parrot & Standard PHP interpreter is driven by the Zend Engine \cite{PHPGeneralInformation, phpPHPHistory}               \\
\rowcolor{white}
Perl                               & perl5 \cite{PerlPerl52024a}                          & N/A (for Perl 5) & Only implementation for Perl 5 \cite{PerlWwwPerl}               \\
\rowcolor{table_gray!50}
Rust                               & rustc \cite{RustlangRust2024a}                       & mrustc, gccrs & Official and only fully functional compiler for Rust \cite{rustlangWhatRustc}               \\
\rowcolor{white}
Go                                 & gc \cite{GolangGo2024}                               & gccgo, gofrontend, TinyGo, GopherJS, yaegi & Official compiler included in Go releases \cite{IntroductionGoCompiler}              \\
\rowcolor{table_gray!50}
Python                             & CPython \cite{PythonCpython2024}                     & PyPy, Stackless Python, MicroPython, CircuitPython, IronPython, Jython & Reference implementation \cite{PythonDeveloperGuide}               \\
\bottomrule
\multicolumn{4}{l}{\small PL: Programming Language, Impl.: Implementation (runtime environment, engine, interpreter, or compiler)}
\end{tabular}
}
\end{table*}

\aptLtoX{\begin{table}
\caption{Versions used in the ReDoS defense measurements.}
\label{tab:engine_versions}
\resizebox{\linewidth}{!}{%
    \small
    \begin{tabular}{p{3.5cm}p{2cm}p{2cm}}
      \toprule
      \textbf{Language} & \textbf{Old Version} & \textbf{New Version} \\
      \midrule
      JavaScript (\textit{Node.js---V8}) & v15.14.0 & v22.2.0 \\
      Ruby (\textit{MRI/CRuby})           & 3.1.6   & 3.3.2   \\
      C\# (\textit{.NET})                & 6.0.420 & 7.0.407 \\
      Java (\textit{OpenJDK})            & 8u342   & 23      \\
      PHP (\textit{Zend Engine})         & 5.6.40  & 8.3.7   \\
      Perl (\textit{perl5})              & 5.8.14  & 5.38.2  \\
      Rust (\textit{rustc})              & 1.12.1  & 1.78.0  \\
      Go (\textit{gc})                   & 1.5.4   & 1.22.4  \\
      Python (\textit{CPython})          & 3.6.15  & 3.12.3  \\
      \bottomrule
    \end{tabular}%
  }
\end{table}}{\begin{table*}
\caption{Versions used in the ReDoS defense measurements.}
\label{tab:engine_versions}
    \small
    \begin{tabular}{p{3.5cm}p{2cm}p{2cm}}
      \toprule
      \textbf{Language} & \textbf{Old Version} & \textbf{New Version} \\
      \midrule
      \rowcolor{table_gray!50}
      JavaScript (\textit{Node.js---V8}) & v15.14.0 & v22.2.0 \\
      Ruby (\textit{MRI/CRuby})           & 3.1.6   & 3.3.2   \\
      \rowcolor{table_gray!50}
      C\# (\textit{.NET})                & 6.0.420 & 7.0.407 \\
      Java (\textit{OpenJDK})            & 8u342   & 23      \\
      \rowcolor{table_gray!50}
      PHP (\textit{Zend Engine})         & 5.6.40  & 8.3.7   \\
      Perl (\textit{perl5})              & 5.8.14  & 5.38.2  \\
      \rowcolor{table_gray!50}
      Rust (\textit{rustc})              & 1.12.1  & 1.78.0  \\
      Go (\textit{gc})                   & 1.5.4   & 1.22.4  \\
      \rowcolor{table_gray!50}
      Python (\textit{CPython})          & 3.6.15  & 3.12.3  \\
      \bottomrule
    \end{tabular}%
\end{table*}}

\section{ReDoS in Practice} \label{sec:appendix-REDOSInPractice}

\noindent

We conducted a comprehensive analysis of NVD CVE data from 2014 to 2023 in order to quantify ReDoS vulnerabilities in practice.
As part of this effort, we utilized the CWE field within the CVE reports.
For each OWASP Top 10 category, we counted the number of CVEs that included a CWE entry associated with that specific OWASP category.

To identify ReDoS CVEs, we employed a systematic two-stage keyword search:

\begin{itemize}[leftmargin=20pt]
    \item \textbf{Stage one keywords:} \texttt{regex}, \texttt{regular expression}, \texttt{regexp}
    \item \textbf{Stage two keywords:} \texttt{algorithm}, \texttt{backtrack}, \texttt{uncontrolled}, \texttt{repetition}, \texttt{repeat}, \texttt{infinite}, \texttt{denial of service}, \texttt{dos}, \texttt{infinite loop}, \texttt{algorithmic complexity}
\end{itemize}

This method effectively identified ReDoS-related CVEs, and our verification of a sample confirmed no false positives.

\section{Works Excluded From the Literature Review}\label{subsec:excluded-works}

Some closely related papers were excluded from the literature review because they do not focus on ReDoS-specific vulnerabilities or their evaluation lacks any ReDoS-specific aspects. These exclusions were necessary to ensure that the review remains precise and relevant to the study of ReDoS.
For instance, \cite{coxRegularExpressionMatching2007, varatalu2025re, shan2017tail} were excluded because they focused on broader DoS vulnerabilities or algorithmic complexity attacks. Other papers were excluded for evaluating performance issues in regex engines without demonstrating their impact on ReDoS vulnerabilities.

Table~\ref{tab:excluded_papers} highlights examples of these excluded works along with the reasons for their exclusion, showcasing the deliberate scope refinement undertaken to focus on ReDoS-specific contributions.

\section{Additional Details of the Engineering Review} \label{sec:appendix-engine-review}

This appendix provides supplementary information to support the engineering review presented in \S\ref{sec:engstudy}. It includes details on the regex engine selection process and the engine versions used in our measurements.

\subsection{Regex Engine Selection}
\label{subsec:regex-engine-selection-summary}

\JD{Place the definition of ``reference implementation'' somewhere in here (maybe the table caption). Put the key term in italics.}

Table~\ref{tab:appendix-engine_selection} summarizes the regex engine implementations reviewed in our ReDoS defenses review (\S\ref{sec:engstudy}). For each programming language considered in our study, we list the selected implementation, notable alternatives, and our rationale. We prioritized (1) engines widely used in production settings (\eg Node.js---V8, used by 98.8\% of websites employing server-side JavaScript~\cite{DistributionWebServersa}); and (2) reference implementations with publicly accessible source code. These are typically cited in official language documentation and serve as the standard for the language (\eg CPython for Python). These choices ensure both real-world relevance and access to implementation details for assessing ReDoS defenses.

\subsection{Versions Used in the Measurements} \label{sec:engine-versions}

To ensure the reproducibility of the experiments described in \S\ref{sec:Engines-Measurements}, Table~\ref{tab:engine_versions} indicates the specific versions of the programming language runtimes and their regex engines that were evaluated.
The \textit{old} versions correspond to earlier releases that predate known or likely ReDoS-relevant mitigations, while the \textit{new} versions refer to more recent releases that incorporate possible ReDoS defenses or performance improvements in their regex engines.


\section{Analyzing Developer Discussions on GitHub ReDoS Issues}
\label{sec:comments-discussion}
In~\S\ref{sec:discussion}, we discussed the idea of ``\textit{Developers push back}''.
We referred to conversations on GitHub and sites of other engineering discourse.

To find discussions about ReDoS vulnerabilities, we used the GitHub REST API to search for issues containing the term \texttt{redos} in their title, body, or comments.
This approach helped us find both open and closed issues across many repositories, covering a range of cases, including those still active and those already resolved. We started with $\sim$46.4K issues and filtered them to include only those with at least two comments, leaving $\sim$7.4K issues. This helped us focus on discussions with more input. Out of these, $\sim$22.6K issues were not opened by bots. $\sim$4K of those are open, and $\sim$18.6K are closed. We manually reviewed 250 recent issues from each set, along with their comments, to find relevant ReDoS discussions.

Table~\ref{tab:redos_comments} showcases 10 examples of comments from selected issues, providing URLs for further exploration. These discussions often reveal that developers do not perceive ReDoS vulnerabilities as relevant or critical, leading to delays or reluctance in addressing the issues. While some developers prioritize security, others downplay the severity of these vulnerabilities, resulting in limited attention or resolution efforts.
However, we acknowledge this aspect of our investigation is incomplete and bears further study.

\JD{Need to know how many issues there were initially and how many survived the filter.}

\begin{table*}[t]
  \centering
  \caption{Selected comments regarding ReDoS and corresponding issue URLs.}
  \label{tab:redos_comments}
  \resizebox{\textwidth}{!}{%
    \begin{tabular}{p{12cm} p{7.75cm}}
      \toprule
      \rowcolor{white}\textbf{Comment} & \textbf{URL} \\
      \midrule
      \rowcolor{table_gray!50} D3 is almost never used to handle user input, so these ReDoS vulnerabilities are generally a huge waste of time.
        & \url{https://github.com/d3/d3/issues/3939} \\
      \rowcolor{white} indeed, but since you'd have to be attacking yourself to trigger the ReDOS in npm, it's not actually a vulnerability here.
        & \url{https://github.com/npm/cli/issues/7902} \\
      \rowcolor{table_gray!50} None of these seem like they could be exploitable in real world scenarios. Please let us know if you disagree.
        & \url{https://github.com/facebook/react-native/issues/46996} \\
      \rowcolor{white} The few times there was an actual vulnerability, it was reported separately, and we released patches as soon as it was possible. You can always report real vulnerabilities here, but please do this if you understand the difference between a real vulnerability and a false positive. For example, a "Regex DDOS attack" can never be a real vulnerability for a development-time tool. If you're not sure, you're welcome to ask in this thread, but please keep it brief and to the point so that the thread doesn't become unreadable.
        & \url{https://github.com/facebook/create-react-app/issues/11174} \\
      \rowcolor{table_gray!50} This kind of low-quality finding was assigned a CVE ID without any significant cross-checking.
        & \url{https://github.com/pytest-dev/py/issues/287} \\
      \rowcolor{white} So, a DoS can be caused by a package that's considered for download and installation (which can run arbitrary code), or by the site that serves such packages? I don't see how this can be exploited without the attacker being able to do much more damage; a fix will mainly silence automated checkers.
        & \url{https://github.com/python/cpython/issues/102202} \\
      \rowcolor{table_gray!50} Most ReDOS vulnerabilities are self-attacks, meaning, not a vulnerability.
        & \url{https://github.com/sarbbottam/eslint-find-rules/issues/349} \\
      \rowcolor{white} Pretty much all complex regexes are vulnerable to ReDos. I’ve migrated some more important ones to use a token parsing approach instead of regexes.
        & \url{https://github.com/nodemailer/mailparser/issues/378} \\
      \rowcolor{table_gray!50} When evaluating whether something is a vulnerability, you have to look at the attack vector and the respective cost of upgrading.
        & \url{https://github.com/facebook/docusaurus/issues/10491} \\
      \rowcolor{white} Anyone using any utils with inputs of arbitrary length runs a performance risk. Even built-ins aren't immune.
        & \url{https://github.com/bestiejs/platform.js/issues/139} \\
      \bottomrule
    \end{tabular}%
  }
\end{table*}

\end{document}